\documentclass[final, 5p, times, twocolumn,authoryear]{elsarticle}
\usepackage[utf8]{inputenc}
\usepackage{booktabs}
\usepackage{url}
\usepackage{todonotes}
\usepackage{afterpage}

\newcommand{\uV}{$\mathrm{\mu V}$}

\journal{ }

\begin{document}

\begin{frontmatter}

\title{Quantity versus Diversity: Influence of Data on Detecting EEG Pathology \\ with Advanced ML Models}

\author[4,1]{Martyna Poziomska}
\ead{martyna.poziomska@fuw.edu.pl}

\author[1,2]{Marian Dovgialo}
\ead{marian.dovgialo@fuw.edu.pl}

\author[1,5]{Przemysław Olbratowski}
\ead{p.olbratowsk@uw.edu.pl}

\author[3]{Paweł Niedbalski}
\ead{pawel.niedbalski@elmiko.pl}

\author[3]{Paweł Ogniewski}
\ead{po_notice@elmiko.pl}

\author[3]{Joanna Zych}
\ead{joanna.zych@elmiko.pl}

\author[1]{Jacek Rogala}
\ead{jacek.rogala@fuw.edu.pl}

\author[1]{Jarosław Żygierewicz\corref{cor1}}
\ead{jaroslaw.zygierewicz@fuw.edu.pl}
\cortext[cor1]{Corresponding author}

\affiliation[1]{organization={Faculty of Physics, University of Warsaw}, 
                              addressline={Pasteura 5},
                              postcode={02-093},
                              city={Warsaw}, 
                              country={Poland}}

\affiliation[2]{organization={Nencki Institute of Experimental Biology, Polish Academy of Sciences}, 
                              addressline={Pasteura 3},
                              postcode={02-093},
                              city={Warsaw}, 
                              country={Poland}}

\affiliation[3]{organization={Elmiko Biosignals sp. z o.o.}, 
                              addressline={Sportowa 3},
                              postcode={05-822},
                              city={Milanówek},
                              country={Poland}}
                              
\affiliation[4]{organization={Automotive Industry Institute, Łukasiewicz Research Network}, 
                              addressline={Jagiellońska 55},
                              postcode={03-301},
                              city={Warsaw},
                              country={Poland}}

\affiliation[5]{organization={WIT Academy}, 
                              addressline={Newelska 6},
                              postcode={01-447},
                              city={Warsaw},
                              country={Poland}}


\begin{abstract}
This study investigates the impact of quantity and diversity of data on the performance of various machine-learning models for detecting general EEG pathology. We utilized an EEG dataset of 2,993 recordings from Temple University Hospital and a dataset of 55,787 recordings from Elmiko Biosignals sp. z o.o. The latter contains data from 39 hospitals and a diverse patient set with varied conditions. Thus, we introduce the Elmiko dataset -- the largest publicly available EEG corpus. Our findings show that small and consistent datasets enable a wide range of models to achieve high accuracy; however, variations in pathological conditions, recording protocols, and labeling standards lead to significant performance degradation. Nonetheless, increasing the number of available recordings improves predictive accuracy and may even compensate for data diversity, particularly in neural networks based on attention mechanism or transformer architecture. A meta-model that combined these networks with a gradient-boosting approach using handcrafted features demonstrated superior performance across varied datasets.
\end{abstract}


\end{frontmatter}

\section{Introduction}

Applications of machine learning (ML) in medical diagnostics are developing rapidly and offer innovative solutions for exploring complex biomedical data. Electroencephalography (EEG), which non-invasively records brain activity, is crucial for diagnosing neurological conditions such as epilepsy \citep{Yang2022eeg}. However, it presents significant challenges for automatic analysis due to its non-linear, non-stationary, and chaotic nature \citep{klonowski2002complexity}. The agreement among human EEG raters is moderate and fluctuates across various neurological disorders \citep{ahrens2021electroencephalography, halford2015inter, nascimento2024expert}, complicating the task further.

\begin{figure*}[!htb]
\centering
\includegraphics[width=0.5\textwidth]{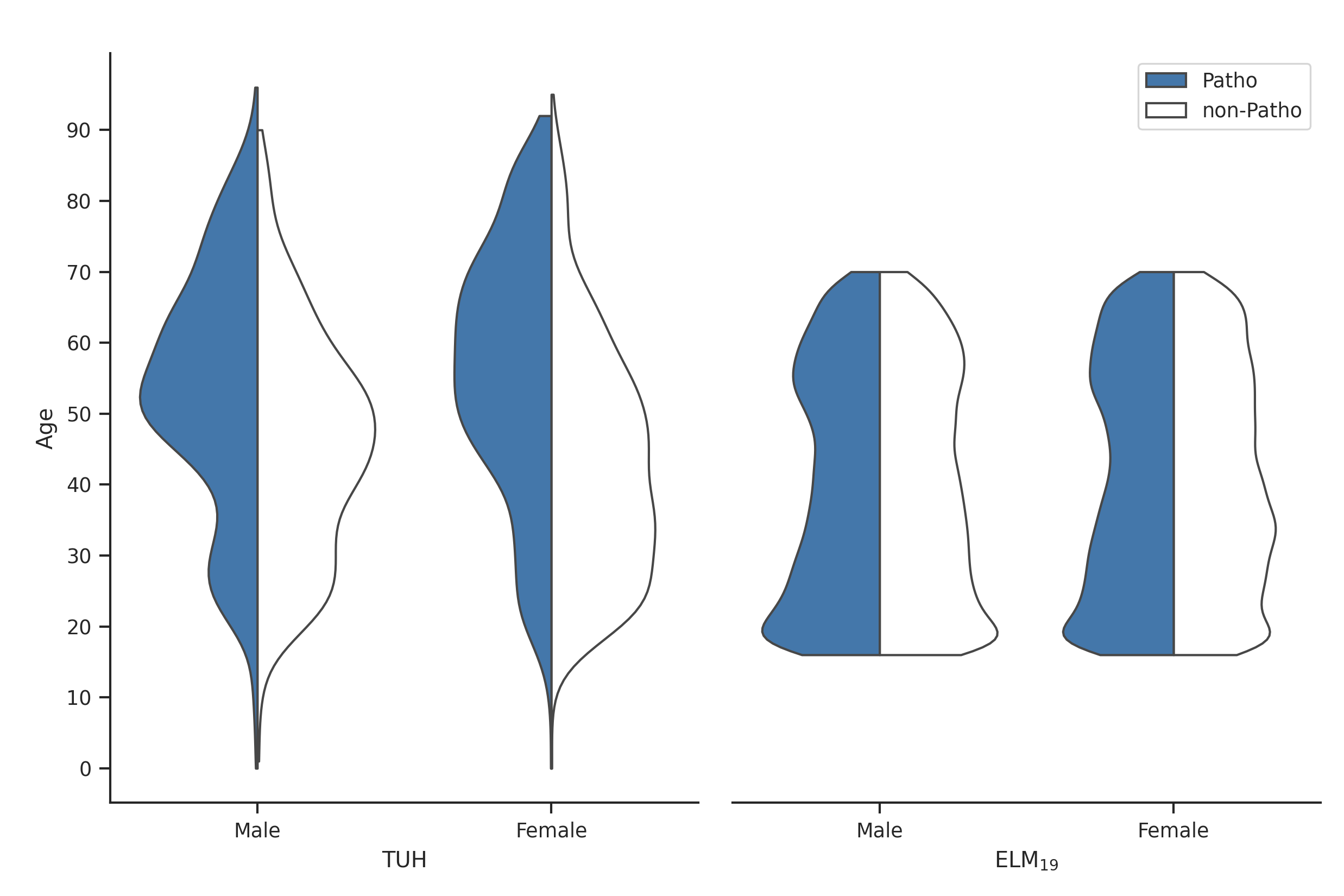}
\caption{Age, sex, and diagnosis distributions in the TUH and ELM$_{19}$ databases.} 
\label{fig:age}
\end{figure*}

Nevertheless, several ML architectures have been proposed to address challenges, such as detecting epileptic seizures or classification of EEG recordings as normal or pathological. The present paper focuses on the latter application, commonly referred to as neuro-screening. Previous research on ML models for general pathology detection includes studies by \citet{TUH_DB}, \citet{schirrmeister2017deep}, \citet{roy2019chrononet}, \citet{leeuwen2019detecting}, \citet{Gemein2020}, \citet{western2021automatic}, \citet{khan2022nmt}, \citet{wilson2022deep}, \citet{kiessner2023extended}, and \citet{kiessner2024reaching}. Review articles by \citet{faust2018deep}, \citet{miotto2018deep}, \citet{craik2019deep}, \citet{roy2019deep}, \citet{merlin2022deep}, and \citet{tibermacine2024riemannian} provide further insights. The proposed architectures include classical methods such as random forest classifiers, support vector machines, Riemannian geometry, and deep neural networks.

ML models generally perform well with homogeneous data. For example, the Temple University Hospital (TUH) corpus has become the de facto benchmark in the field. It comprises approximately 3,000 labeled clinical recordings, making it the largest publicly available dataset for detecting EEG pathology to date. All recordings originate from a single institution and were labeled by a dedicated team of clinicians who exhibited a high inter-rater agreement. With over 87\% of the abnormal recordings indicating epilepsy \citep{obeid2016temple}, the classification task primarily focuses on distinguishing epilepsy from normal EEG patterns.

Despite demonstrating high accuracy with minimal electrode use \citep{yildirim2020deep} and achieving 87.7\% accuracy in recent studies utilizing wavelet feature extraction or gradient boosting \citep{albaqami2021automatic}, models have not achieved significantly better results on this seemingly straightforward task. \citet{kiessner2024reaching} note that predictive accuracy tends to plateau at around 87\%, attributing this limitation to the inherent noise in clinical labeling. They also indicate that although expanding the amount of data can enhance model capabilities, the benefits are not uniform across all model architectures. Simpler models may experience a plateau in accuracy as data volume increases, whereas more complex models, particularly neural networks with residual blocks, continue to improve.

Efforts to automatically detect EEG abnormalities have mostly focused on training ML models using signals from a single hospital, with TUH being a prominent example. Although a resulting model may achieve remarkably high accuracy for that particular hospital, its generalizability to other institutions is often poor. A similar effect has been reported by \citet{ayodele2020supervised} in the context of automatic detection of EEG seizures. Data from one institution may be biased regarding patient conditions, and the corresponding model can fail even when the same institution admits less typical patients. A change in EEG hardware or the introduction of new technicians is also likely to invalidate the current model. Finally, a single hospital may not provide enough recordings for reliable training.
 
The need to incorporate data from multiple sources has been widely recognized in other medical domains. Examples include predicting the risk of mortality in cardiac surgery \citep{roques1999risk}, the risk of intracranial aneurysm rupture \citep{greving2014development}, the risk of coronary artery disease \citep{aerts2017pooled}, and the risk of major bleeding in patients with noncardioembolic stroke \citep{hilkens2017predicting}. Training on data from multiple institutions is often referred to as multi-cohort learning. \citet{schinkel2023embracing} embrace the intuitive idea that training on multiple cohorts yields models that better generalize to new cases from hospitals not included in the training set.

However, multi-cohort learning comes with its own difficulties. Clinical data from different hospitals can be heterogeneous or diverse in terms of patient conditions, recording protocols, and labeling standards \citep{tayebi2023enhancing}. Such data occupy different regions of the feature space and introduce greater labeling noise, which can significantly deteriorate model performance, even for hospitals included in the training set. In a recent study of phenotype predictions, \citet{gao2023batch} observed a decline in accuracy across all considered statistical methods as population heterogeneity increased. There is certainly a trade-off between pros and cons. Nevertheless, attempts to develop an approach applicable to virtually all EEG recordings are undoubtedly worth the effort for scientific purposes, for the benefit of patients, and also for commercial reasons.

Given these observations, the present study investigates the performance of simple and advanced ML models on EEG data of varying volumes and diversity. We use the relatively small, homogeneous TUH dataset and a large, heterogeneous dataset provided by Elmiko Biosignals sp. z o.o., a Polish manufacturer of EEG hardware and software. The latter dataset includes recordings from multiple hospitals, evaluated by many independent experts, and reflects diverse patient conditions. Our models include both non-neural approaches that utilize handcrafted features and neural networks that extract features directly from the raw EEG signal. Some of these networks use the attention mechanism or the transformer architecture. By examining how different models respond to variations in data quantity and heterogeneity, we aim to identify the most effective approaches for developing robust EEG classifiers capable of generalizing across diverse clinical settings and patient conditions.

Our findings contribute to the broader discussion on the role of data diversity in ML, particularly within the medical field. The insights gained from this study are expected to inform the design of future ML architectures and guide the collection of medical datasets, ultimately advancing the development of more reliable and generalizable diagnostic tools.

\section{Data}
\label{sect:data}

\begin{table*}[!htb]
\caption{Descriptive statistics of the datasets. \label{tab:databeses}}
\centering
\begin{tabular}{llllllll}
\toprule
             & TUH   & SZC   & ELM$_1$ & ELM$_2$ & ELM$_4$ & ELM$_8$ & ELM$_{19}$ \\
\midrule
Recordings   & 2,993 & 2,993 & 2,993   & 5,986   & 11,972  & 23,944  & 55,787     \\
Females      & 1,595 & 1,533 & 1,630   & 3,205   & 6,334   & 12,675  & 29,525     \\
             & 53\%  & 51\%  & 54\%    & 54\%    & 53\%    & 53\%    & 53\%       \\
Pathological & 1,472 & 1,497 & 1,561   & 3,119   & 6,238   & 12,479  & 29,074     \\
cases        & 49\%  & 50\%  & 52\%    & 52\%    & 52\%    & 52\%    & 52\%       \\
\bottomrule
\end{tabular}
\end{table*}

In the current study, we contrast two distinct sources of data. The first is the publicly available TUH Abnormal EEG Corpus version 2.0.0, which is part of a larger database \citep{TUH_DB}. It contains 2,993 recordings, of which 1,595 (53.3\%) are from females. A total of 1,472 recordings (49.2\%) are suggestive of pathology, with the vast majority (87\%) classified as epileptic. The remaining 1,521 recordings are classified as normal. Patient ages range from 0 to over 90 years. We refer to this database as TUH. The age, sex, and pathology distributions in TUH are depicted in Fig. \ref{fig:age}.

The second database was collected by Elmiko Biosignals sp. z o.o.\footnote{\url{https://www.elmiko.pl}} and originally contained over 80,000 recordings, each accompanied by a description in Polish. This description includes the final diagnosis based on signal examination by an EEG expert. We manually reviewed all descriptions and excluded cases where the diagnosis was missing or inconclusive. Specifically, we discarded descriptions indicating the absence of epileptic abnormalities without commenting on other possible disorders, as well as those emphasizing that the observed abnormalities were weak \citep{benbadis2010tragedy}. Recordings with descriptions citing significant or abundant artifacts were also excluded. We retained only those recordings that unequivocally reported the presence or absence of abnormalities or explicitly stated that the signal was normal, labeling them accordingly as pathological or normal.

After additional exclusions due to the preprocessing described in Section \ref{sect:preproc}, the database contains 55,787 recordings, of which 29,525 (52.9\%) are from females. Patient ages vary from 16 to 70. A total of 29,074 recordings (52.2\%) are described by clinicians as indicative of pathology. Among these, only about 26\% are recognized as epileptic or exhibiting seizures. The remaining 26,713 recordings are identified as normal. Importantly for this study, the data were collected from 39 different hospitals, making it strongly heterogeneous in terms of recording procedures and labeling standards. This database will be referred to as ELM$_{19}$ because it is roughly 19 times the size of TUH. The age, sex, and pathology distributions in the ELM$_{19}$ database are shown in Fig. \ref{fig:age}. Note that ELM$_{19}$ and TUH have similar proportions of sex and pathology.

We also examine several subsets of the ELM$_{19}$ database. The first subset, named SZC, is designed to closely match the TUH database in terms of size, the proportion of female participants, and the ratio of normal to pathological recordings. Notably, it originates from a single hospital, making it significantly more homogeneous compared to the entire ELM$_{19}$ database, while still covering a broader range of pathological conditions than TUH. The next five subsets are stratified based on hospital, sex, and pathology, differing only in size. They are labeled ELM$_1$, ELM$_2$, ELM$_4$, ELM$_8$, and ELM$_{19}$, with the subscript indicating how many times each subset is larger than TUH. The last subset is the full ELM$_{19}$ database, which is 18.64 times the size of TUH. Table \ref{tab:databeses} provides descriptive statistics for all datasets. While we use TUH and the complete ELM$_{19}$ to examine the effects of heterogeneity, the different ELM$_{19}$ subsets enable us to assess the impact of varying data sizes.

EEG signals in both databases were recorded using the standard 10-20 system with 19 channels denoted as follows: Fp1, Fp2, F7, F3, Fz, F4, F8, T3, C3, Cz, C4, T4, T5, P3, Pz, P4, T6, O1, and O2. They were initially sampled at various frequencies ranging from 200 to 500\,Hz. The TUH and ELM$_{19}$ signals were recorded in the United States and Europe, where the power-line frequency is 60 and 50\,Hz, respectively.

While the homogeneous TUH data have been openly available for several years, we are now sharing the heterogeneous ELM$_{19}$ database for research purposes, making it the largest publicly accessible EEG corpus for pathology detection. Please refer to the data availability declaration at the end for details. The recordings in both databases are originally stored as separate files in the European Data Format\footnote{\url{https://www.edfplus.info}} (EDF) \citep{kemp1992simple}, which can be easily read using, e.g., open-access \texttt{Python} libraries.

\section{Methods}

\subsection{Preprocessing}
\label{sect:preproc}

We preprocessed raw signals from both databases using the \texttt{MNE}\footnote{\url{https://mne.tools}} package for EEG analysis \citep{GramfortEtAl2013a}. To remove power-line noise, a notch filter with quality factor $Q = 5$ and frequencies of 60 and 50\,Hz was applied for the TUH and ELM$_{19}$ databases, respectively. Butterworth filters were then used to eliminate frequencies often contaminated by slow drift and muscle artifacts. We applied a high-pass filter with a cutoff at 0.1\,Hz and less than 1\,dB passband ripple above 0.5\,Hz, and a low-pass filter with a cutoff at 40\,Hz, providing at least 20\,dB attenuation above 50\,Hz, and less than 1\,dB passband ripple. All signals were then resampled to 100\,Hz and re-referenced to a common average reference.

\begin{figure*}[!htb]
\centering
\includegraphics[width=0.75\textwidth]{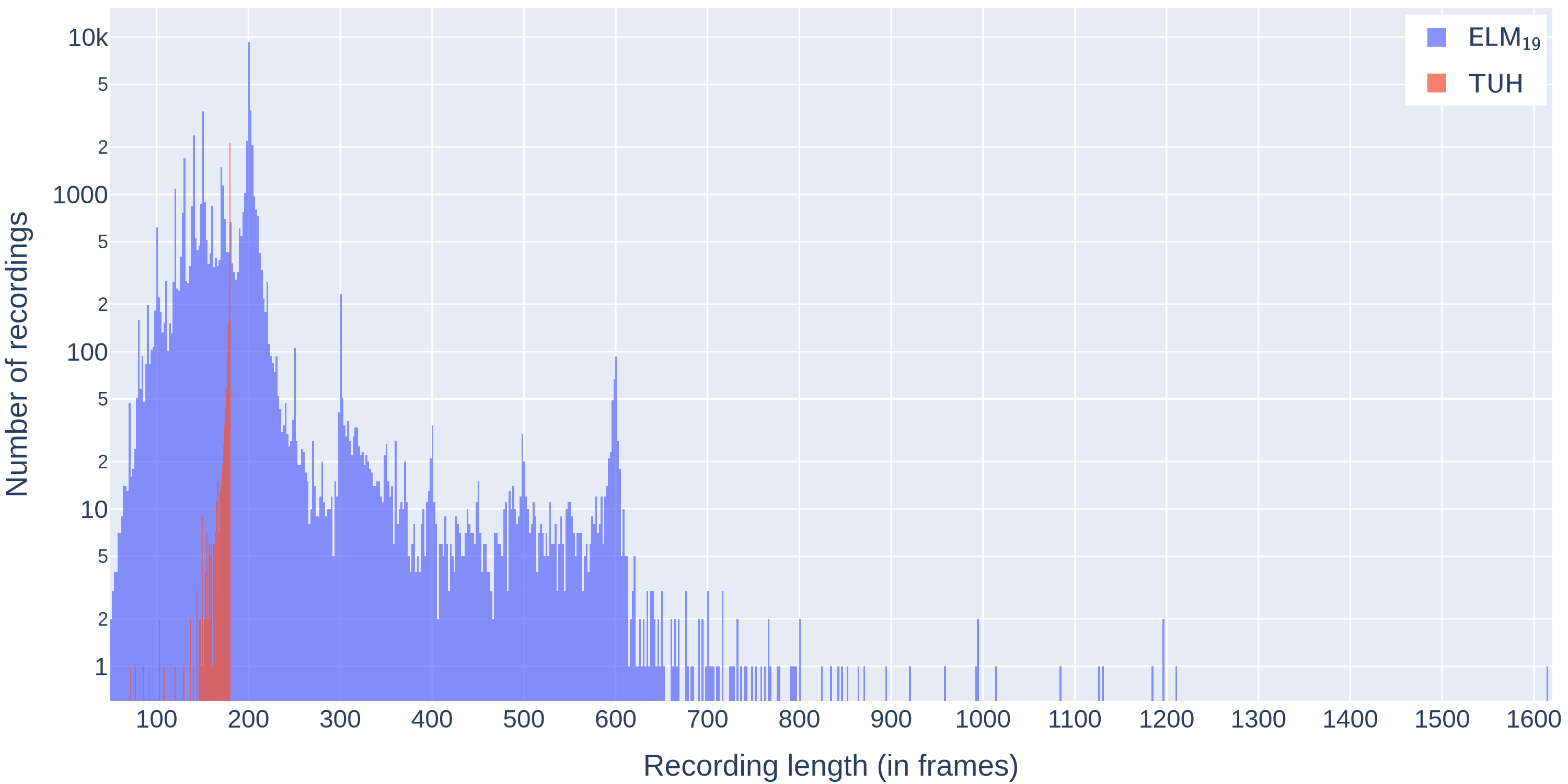}
\caption{Distribution of recording lengths in the TUH and ELM$_{19}$ databases after data preprocessing. The length is expressed as the number of 6-second frames.} 
\label{fig:frames}
\end{figure*}

Following \citet{Gemein2020}, signals were split into adjacent 6-second frames, each containing 600 EEG samples. We checked that slicing frames of 10 or 20 seconds did not improve the results further. Frames containing flat-line channels or voltages exceeding 800\,\uV\, were discarded as artifacts. Recordings with fewer than 50 valid frames remaining after artifact removal were excluded from further analysis. Although the same procedure was applied to both databases, it did not reject any recordings from TUH. Note that all the numbers given in Section \ref{sect:data} pertain to recordings remaining after all the described exclusions and that were indeed involved in the training and testing of our ML models. Fig. \ref{fig:frames} shows the resulting distributions of the recording lengths in both datasets. The histograms highlight the greater diversity of ELM$_{19}$ in this respect as well.

To comply with the aggregation methods used by some of our models (see Section \ref{sect:net}), we label normal recordings as 1 and pathological recordings as 0, contrary to the convention commonly adopted in medicine. Consequently, our models technically output the probability of normality rather than pathology. Because these two quantities are trivially related, we sometimes refer to the models as returning the probability of pathology, which may be more appealing to the reader.

\subsection{Feature extraction}
\label{sect:feature}

We study a variety of neural networks alongside a group of non-neural models, referred to as classical models. The neural networks take the preprocessed signal frames directly and automatically extract features useful for detecting EEG pathology. In contrast, the classical models rely on handcrafted features based on time and frequency.

Time-domain covariance matrices between EEG channels form a Riemannian manifold and their projections onto the tangent space carry valuable information about the EEG signal \citep{congedo2013new, tibermacine2024riemannian}. For every 6-second frame, we calculate the covariance matrix, average those matrices over frames in each recording using a Riemannian metric \citep{Maher2005}, and then map them to the tangent space \citep{Lotte2018}. The projections yield vectors of 190 features, corresponding to the number of elements in the upper or lower triangles of the original symmetric matrices. We computed the time-domain features end-to-end from the preprocessed signals using the \texttt{pyRiemann}\footnote{\url{https://pyriemann.readthedocs.io}} library.

To construct the frequency-based features, we consider 14 frequency bands ($f_b$): \{[0.5, 2], [1, 3], [2, 4], [3, 6], [4, 8], [6, 10], [8, 13], [10, 15], [13, 18], [15, 21], [18, 24], [21, 27], [24, 30], [27, 40]\}\,Hz. These bands cover the standard EEG frequency bands of $\delta$, $\theta$, $\alpha$, low $\beta$, and high $\beta$.

The most basic features in the frequency domain are the power spectral densities corresponding to the specified bands, denoted as $S_{xx}(f_b)$ where $x$ is the channel index. The densities are estimated separately for each frame using the multitaper method \citep{thomson1982spectrum} and then normalized so that the sum over channels and frequency bands equals one in each frame. These estimates are aggregated over frames in each recording using the median operation, following \citet{Gemein2020}. This process results in 266 power features per recording, which were extracted from the EEG signals using the \texttt{MNE} library.

Cross-spectral densities, $S_{xy}(f_b)$, are also calculated for each 6-second frame and are used to estimate the band-wise coherences between EEG channels in that frame according to the following formula:
\begin{equation}
C_{xy}(f_b) = \frac{| S_{xy}(f_b) |}{\sqrt{ S_{xx}(f_b) \cdot S_{yy}(f_b)}}.
\label{eq:coh}
\end{equation}
These coherences are aggregated over frames in each recording using the median. We also considered all other connectivity estimates provided by the \texttt{MNE} library and found the quantities (\ref{eq:coh}) to give the best results. Because the coherence matrix is symmetric with ones on the diagonal, only its sub-diagonal elements are further used. This results in 2,394 coherence features per recording, forming the second group of frequency-based features.

Different classical models considered in this study use various subsets of the handcrafted features, as described in the next section.

\subsection{Models}
\label{sect:models}

We explore a variety of ML models, ranging from classical methods to advanced neural networks. Two of these models, siNet\footnote{Our siNet is equivalent to the model known as BD-EEGNet in the original paper by \citet{Gemein2020}.} and RG, are closely reproduced following \citet{Gemein2020}, while RF is loosely inspired by that paper. RF is based on the random-forest classifier but uses a smaller and simpler set of input features. These three models serve as a state-of-the-art reference, while the other eight models proposed in this paper are their extensions.

\subsubsection{Classical models}

The random-forest\footnote{The RF model is implemented as an instance of the {\tt RandomForestClassifier} class with hyperparameters set according to \citet{Gemein2020}: {\tt criterion='entropy', bootstrap=False, max\_features='sqrt', min\_samples\_leaf=2, min\_samples\_split=2, max\_depth=90,} and {\tt n\_estimators=1600}.} (RF) classifier is implemented using the \texttt{scikit-learn}\footnote{\url{https://scikit-learn.org}} package \citep{scikit-learn}. It uses only the handcrafted power and coherence features, totaling 2,660 features per recording.

The model based on Riemannian geometry (RG) takes the 190 time-domain features as input and passes them to a support-vector classifier\footnote{The RG model is implemented as an instance of the {\tt SVC} class with the following hyperparameters: {\tt kernel='rbf', C=10, probability=True,} and {\tt gamma='auto'}} (SVC) with a radial basis function (RBF) kernel, implemented using the \texttt{scikit-learn} library.

A model we call the gradient-boosted ensemble (GBE) is implemented using the \texttt{CatBoost}\footnote{\url{https://catboost.ai}} library \citep{catboost}. It incorporates all 2,850 handcrafted features and passes them to an ensemble of 30 gradient-boost classifiers\footnote{A single GBE classifier is implemented as an instance of the {\tt CatBoostClassifier} class with the following hyperparameters:  {\tt objective='Logloss',}  {\tt boosting\_type='Plain',} {\tt bootstrap\_type='MVS',} {\tt eval\_metric='AUC',} {\tt iterations=700,} {\tt learning\_rate=0.085195,} {\tt depth=6.0,} {\tt l2\_leaf\_reg=1.1030,} and {\tt colsample\_bylevel=0.019947}.}. The final probability of EEG pathology is obtained by averaging the probabilities predicted by the 30 members of the ensemble. Each member is trained separately and yields a slightly different classifier because the learning algorithm involves stochastic factors.

Hyperparameters of the GBE model were optimized using tree-structured Parzen estimators (TPE) \citep{bergstra2013making} from the {\tt hyperopt}\footnote{\url{https://hyperopt.github.io}} library with respect to the area under the receiver operating characteristic curve (AUC) values from a 5-fold cross-validation on the TUH dataset. Although the hyperparameters fitted to ELM$_{19}$ are slightly different, they do not significantly alter the resulting AUC scores.

\subsubsection{Neural networks}
\label{sect:net}

We implement all our neural models within the \texttt{PyTorch}\footnote{\url{https://pytorch.org}} framework and train them on a single PC equipped with a gaming-class GPU with 12\,GB of memory. A single training session does not significantly exceed one day, even for the most complex transformer models fitted to the largest ELM$_{19}$ dataset.

\begin{figure}[!htb]
\centering
\includegraphics{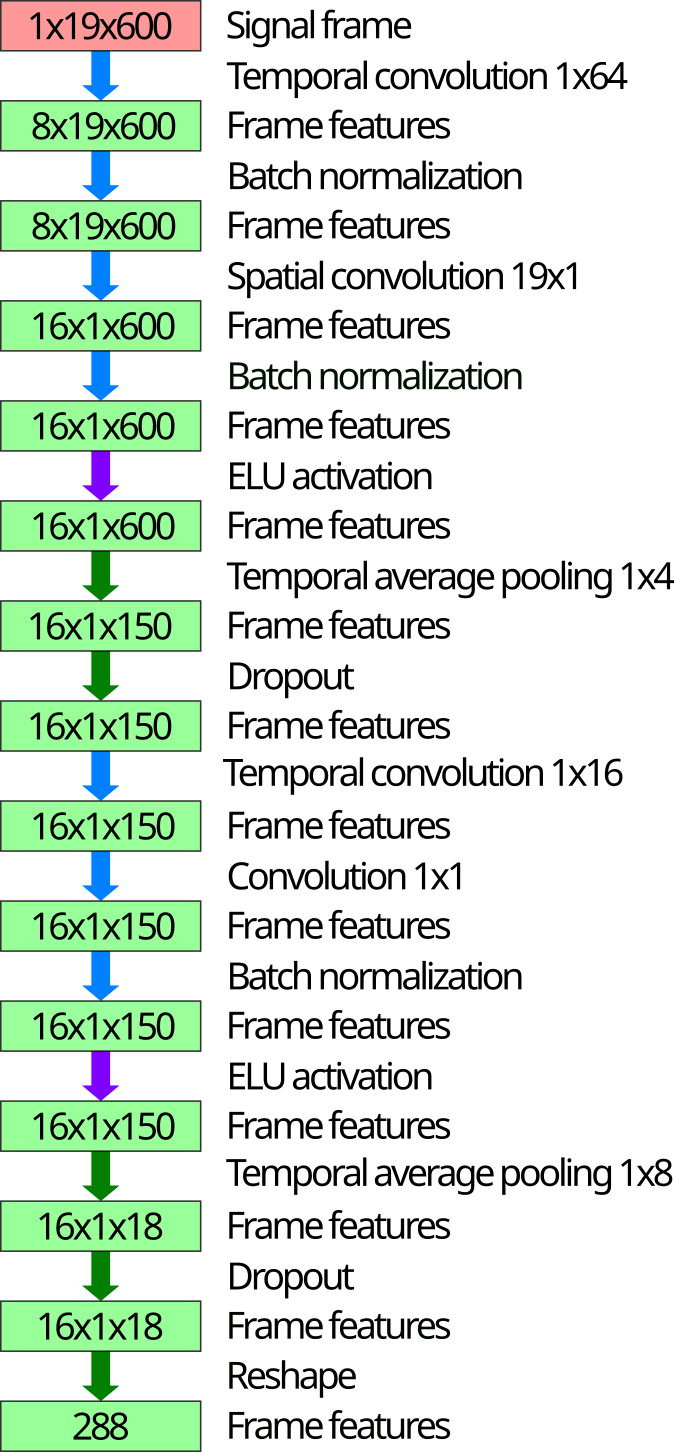}
\caption{Architecture of the EEGNet frame encoder, which processes a 6-second frame of the EEG signal and outputs an encoding as a vector of 288 features. This encoder is reimplemented from \citet{Gemein2020} and \citet{lawhern2018eegnet} and has 1,408 parameters. Arrows represent layers and operations, while boxes indicate tensor shapes. Refer to Fig. \ref{fig:colors} for an explanation of the color code.}
\label{fig:eegnet}
\end{figure}

All our neural networks use a frame encoder that follows the EEGNet architecture depicted in Fig. \ref{fig:eegnet}. This encoder is reimplemented based on the Braindecode EEGNet described by \citet{Gemein2020}, which itself is a reimplementation of the EEGNet originally introduced by \citet{lawhern2018eegnet}. The encoder is a fully convolutional network that performs separate temporal and spatial convolutions. It has 1,408 parameters and constitutes the first part of all our neural models, as described below. We tested other encoders, such as the Deep4 architecture discussed by \citet{Gemein2020}, but found no advantages over EEGNet.

\begin{figure}[!htb]
\centering
\includegraphics{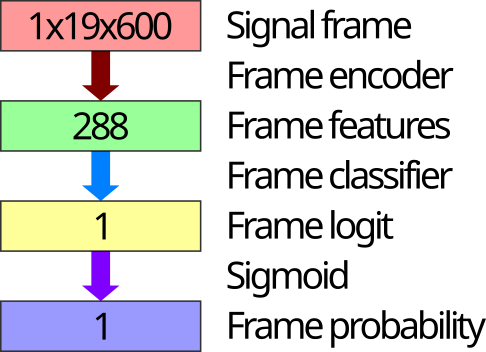}
\caption{Architecture used to train the siNet classifier and encoder, as well as to pretrain the encoder on single frames for later use with the miNet, MINet, and TransNet models. This network has 1,697 parameters.}
\label{fig:pretrain}
\end{figure}

\begin{figure}[!htb]
\centering
\includegraphics{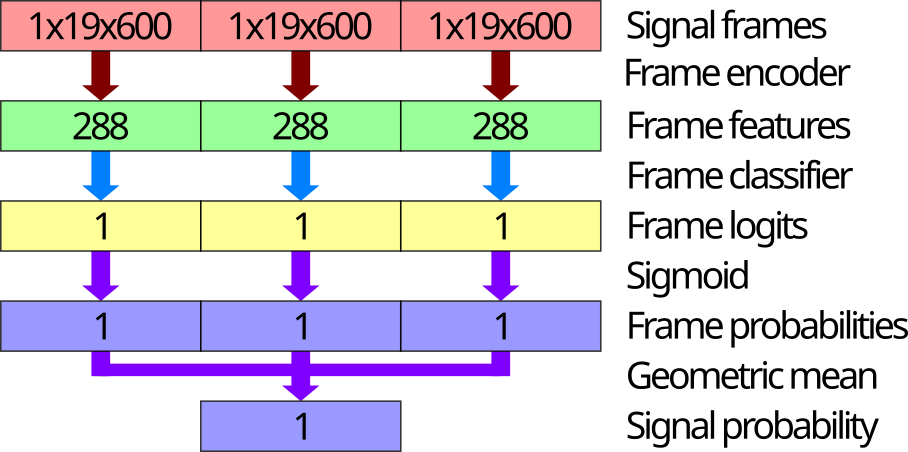}
\caption{Architecture used by the siNet model for prediction and by the miNet model for both prediction and training. This network has 1,697 parameters. Although three frames are shown for illustration, the model can process an entire recording composed of an arbitrary number of frames.}
\label{fig:mininet}
\end{figure}

The simplest way to utilize the frame encoder is shown in Fig. \ref{fig:pretrain}. The encoder is attached to a single fully connected layer that serves as a frame classifier. During the training process, the network is fed individual 6-second frames, and the target label for each frame corresponds to the label of the entire recording from which the frame originates. Predictions are made using the same classifier and encoder arranged as shown in Fig. \ref{fig:mininet}. Each frame of the examined recording is passed through the encoder, classifier, and sigmoid function to yield the probability of the signal in that frame being normal. These probabilities are then aggregated over all frames in the recording using the geometric mean to produce the final probability of normality for the entire recording. We refer to this basic model as siNet because it is trained on single frames.

We tested other methods of aggregating probabilities, such as the arithmetic average, minimum or maximum, but found the geometric mean to give the best results. However, it has one particular property: even if merely one frame in a recording yields zero individual probability, the aggregated probability also becomes zero. On the other hand, we want our models to have high recall and detect pathology even if a single frame exhibits abnormal symptoms. To reconcile these points, we make our models output the probability of the recording being normal rather than pathological; therefore, we adopt the unconventional labeling introduced in Section \ref{sect:preproc}.

The siNet has 1,697 parameters, both in prediction and training, with the classification layer alone accounting for 289 parameters. In fact, the latter layer is used by all our network models as a classifier of either individual frames or entire recordings, as specified below.

Except for the siNet, all our networks are trained in two distinct ways. The first method, which we call naive, involves training the encoder from scratch along with the entire network. In the second approach, we first pretrain the encoder in a process identical to training the siNet, but the resulting classifier is discarded. The pretrained encoder is then attached to the other parts of the network and the entire network undergoes the main training. We denote the pretrained and naive versions of a model by suffixes P and N, respectively, such as miNetN or TransNetP, as described below.

The siNet is trained by assigning the label of the full recording to each of its frames. This is not fully justified because a pathology may not manifest itself in every single frame, and passing normal-like frames labeled as abnormal may confuse the model. To address this simplification, we resort to multiple-instance learning (MIL), a term first coined by \citet{dietterich1997solving} and later discussed by several authors, including \citet{andrews2002support} and \citet{wang2018}. In the current context, the main idea is to train a model on all the frames in a recording simultaneously. Roughly speaking, this can be done using some form of aggregation before evaluating the loss. Note that the way of training our classical models can be viewed as a realization of the MIL paradigm because the handcrafted features of individual frames are aggregated using the median operation or the Riemannian metric.

In the realm of neural networks, successive efforts by \citet{ramon2000multi}, \citet{zhou2002neural}, and \citet{wang2018} led to a simple MIL architecture called miNet. It is formally identical to the siNet shown in Fig. \ref{fig:mininet} and the only difference is that the configuration from Fig. \ref{fig:mininet} is used for both prediction and training. In the training process, the network takes all the frames in a recording simultaneously and produces a single probability, which is then compared against the human label for the entire recording. This is possible because the architecture in Fig. \ref{fig:mininet} is actually a single differentiable network that can process an arbitrary number of frames in one pass, without explicit loops. The encoder can be trained with the entire network from scratch or can be pretrained on single frames.

\begin{figure}[!htb]
\centering
\includegraphics{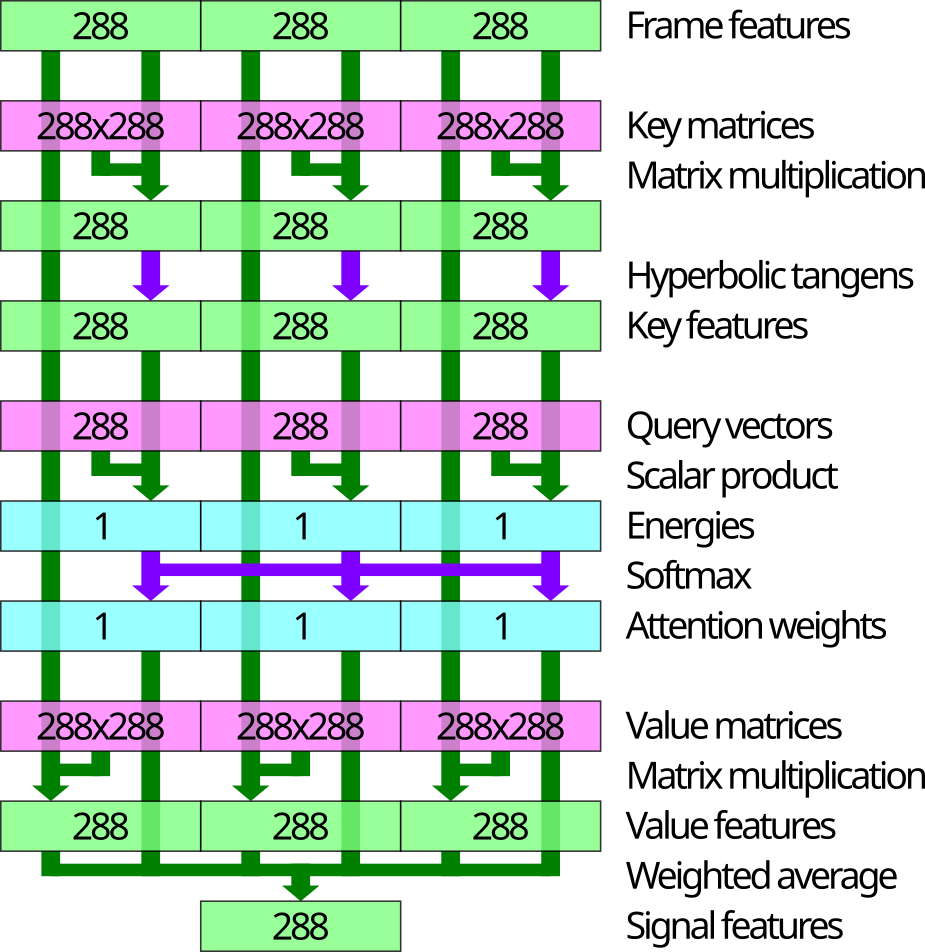}
\caption{Architecture of the attention module. It follows the non-gated attention approach of \citet{ilse2018attention}, while the concepts of query, key, and value are borrowed from the transformer paper by \citet{vaswani2017attention}. This module has 166,176 parameters.}
\label{fig:attention}
\end{figure}

Although miNet is trained on entire recordings, aggregating probabilities still has its limitations. A better solution is to aggregate features, which results in the MINet architecture proposed by \citet{wang2018}. \citet{ilse2018attention} further improved the idea by introducing the attention mechanism, originally considered in the context of machine translation and recurrent networks \citep{bahdanau2014neural, raffel2015feed}. Our variant of the attention module is depicted in Fig.~\ref{fig:attention}. It essentially follows the non-gated attention of \citet{ilse2018attention} but also employs the concepts of queries, keys, and values, borrowed from the transformer architecture \citep{vaswani2017attention}. The attention module takes encoded features from all the frames in a recording and assigns a weight to each frame. Frames deemed more important for detecting pathology are expected to receive higher weights. These weights are then used to calculate a weighted average of the individual frame features, resulting in mean features that represent the entire recording. The attention module has 166,176 parameters.

\begin{figure}[!htb]
\centering
\includegraphics{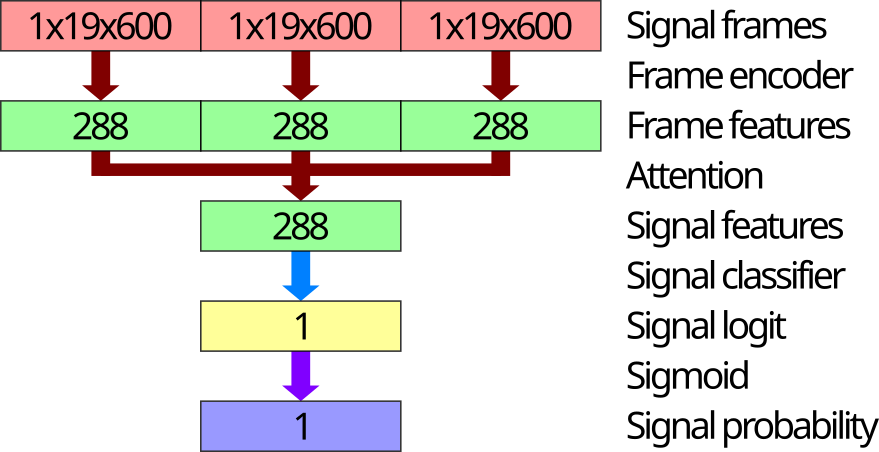}
\caption{Architecture of the MINet model, which has 167,873 parameters.}
\label{fig:maxinet}
\end{figure}

Our MINet model is depicted in Fig. \ref{fig:maxinet}. The network takes an entire recording and passes it through the frame encoder to produce encodings or features for each frame separately. These features then feed the attention module which aggregates them into a vector of mean features that encode the entire recording. These are finally passed to the classifier layer. The same architecture is used for both prediction and training, with the training process being identical to that of miNet. The MINet network has 167,873 parameters. Again, the encoder can be pretrained or not.

\begin{figure}[!htb]
\centering
\includegraphics{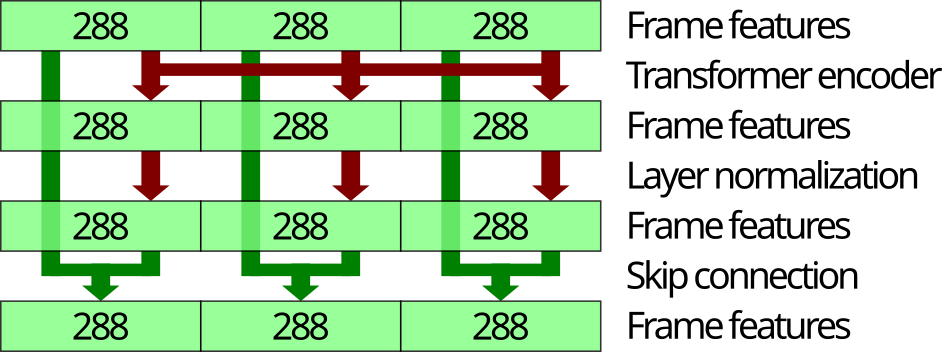}
\caption{Architecture of our custom transformer block, which is the standard transformer encoder layer wrapped with additional layer normalization and a skip connection. This block has 1,516,640 parameters and is utilized in the TransNet model depicted in Fig. \ref{fig:transnet}.}
\label{fig:transformer}
\end{figure}

\begin{figure}[!htb]
\centering
\includegraphics{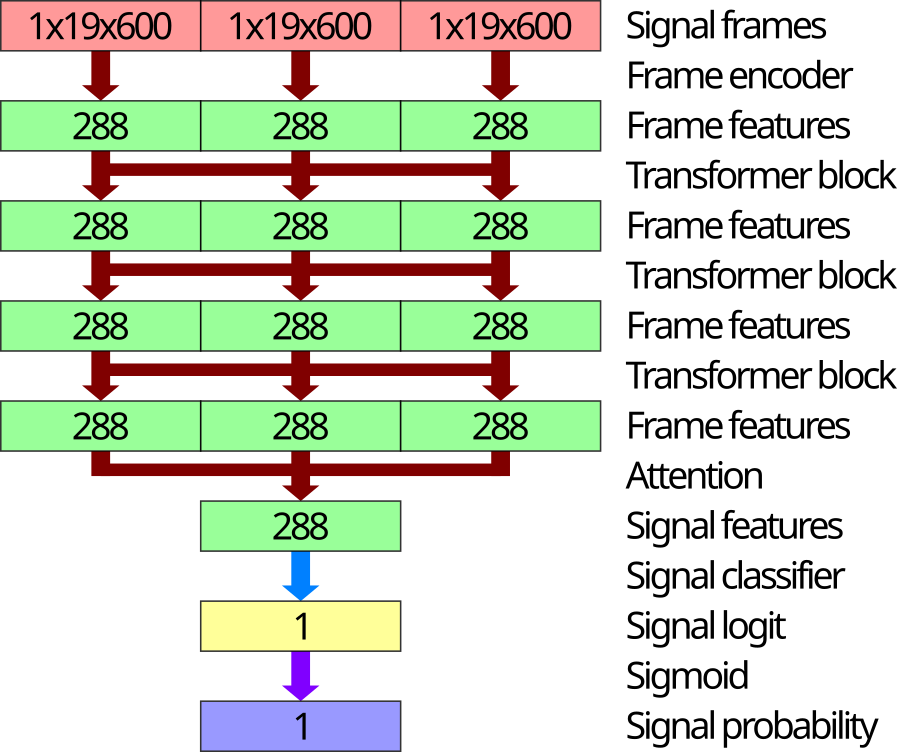}
\caption{Architecture of the TransNet model. The transformer block refers to the custom architecture presented in Fig. \ref{fig:transformer}. The model has 4,717,793 parameters.}
\label{fig:transnet}
\end{figure}

The attention module of MINet effectively selects frames that carry the most information but it cannot account for relations between frames, and, thus, between potentially distant regions of the recording. Such relations can be captured by self-attention, which is the main building block of the transformer architecture \citep{vaswani2017attention}. We utilize only the transformer encoder layer as introduced in the seminal paper and implemented in \texttt{PyTorch} as a predefined class\footnote{The transformer encoder layer is implemented as an instance of the \texttt{TransformerEncoderLayer} class with the following arguments: {\tt d\_model=288, nhead=8,} and {\tt batch\_first=True}.}, with the default number of 8 self-attention heads. However, we found that additional layer normalizations and skip connections slightly improve the results so we employ a custom transformer block depicted in Fig. \ref{fig:transformer}. It has 1,516,640 parameters. We stack three such blocks sequentially. Although it seems sufficient to aggregate the transformer output features using the bare arithmetic mean, we managed to improve performance by using the attention module from Fig. \ref{fig:attention} instead, as in MINet. Our final transformer model is depicted in Fig. \ref{fig:transnet}, and we call it TransNet. It has 4,717,793 parameters.

To recapitulate, we consider 7 neural models, all of which employ the EEGNet as their frame encoder, and we denote them as TransNetP, TransNetN, MINetP, MINetN, miNetP, miNetN, and siNet. They are trained in several ways but for prediction, they all take an arbitrarily long recording split into 6-second frames and return a single probability of pathology. The parameter counts are summarized in Table \ref{tab:parameters}.

\begin{table*}[!htb]
\caption{Parameter counts in our neural models. The transformer column refers to the custom transformer block shown in Fig. \ref{fig:transformer}. \label{tab:parameters}}
\centering
\begin{tabular}{lrrrrr}
\toprule
         & Encoder & Transformer         & Attention & Classifier & Total     \\
\midrule
siNet    & 1,408    &                    &           & 289        & 1,697     \\
miNet    & 1,408    &                    &           & 289        & 1,697     \\
MINet    & 1,408    &                    & 166,176   & 289        & 167,873   \\
TransNet & 1,408    & 3$\times$1,516,640 & 166,176   & 289        & 4,717,793 \\
\bottomrule
\end{tabular}
\end{table*}

\subsubsection{Meta-model}

Different models may exploit different features of the EEG signal. This suggests that an amalgamation of sufficiently distinct models could utilize a richer set of signal traits and, therefore, perform better. Presumably, the handcrafted features of Section \ref{sect:feature} are significantly different from those extracted automatically by the neural networks. Thus, the mixture should include both neural and classical models. After some testing, we decided to combine the classical GBE with the neural MINetP and TransNetP, which are our best-performing single models. Various methods exist for blending several models into one, such as adaptive boosting; however, we empirically chose basic logistic regression. For a given recording, the three probabilities of pathology returned by the component models are fed to the logistic regression as input features and the regression outputs its own final probability. We refer to this solution as a meta-model. The \texttt{scikit-learn} implementation of logistic regression\footnote{The logistic regression model is implemented as an instance of the {\tt LogisticRegression} class with the following hyperparameters: {\tt C=7.9059} and {\tt max\_iter=4,000}.} was used and optimized with the TPE algorithm from the {\tt hyperopt} library, as was done for the GBE.

\subsection{Training, validation, and testing}

Metrics based on the confusion matrix, mostly accuracy (ACC), are frequently used to assess classification quality. However, following criticism from both the medical \citep{beck1986use, bellazzi1998intelligent} and ML \citep{provost1998case, langley2000crafting} communities, we abandon these metrics in favor of the AUC. Among its advantages, the AUC metric is independent of the arbitrary threshold on the probabilities returned by a classification model. We use ACC only for comparison with the results from other articles.

For training the neural models, we minimize the standard binary cross-entropy using the RAdam optimizer \citep{liu2019variance} with the default settings. According to its authors, the optimizer corrects the variance of the adaptive learning rate so that it is consistent, thus eliminating the need for warm-up heuristics.

Each subset of our data -- ELM$_{19}$, ELM$_8$, ELM$_4$, ELM$_2$, ELM$_1$, SZC, and TUH -- has been randomly divided into 6 equal folds, stratified by pathology, sex, and, for subsets of ELM$_{19}$, also by hospitals. For each subset, we carry out a 6-fold cross-validation. In each of its 6 steps, a different fold is selected for testing and a different one for validation, while the remaining four are left as a training set. The cross-entropy is minimized on those 4 training folds, and the AUC value is monitored on the validation fold. One hundred and fifty epochs are performed for the MIL models, while 50 are performed for the siNet and for pretraining the EEGNet encoder. To prevent overfitting, the epoch yielding the highest AUC on the validation fold is chosen as the final one. The resulting model is then evaluated on the test fold to calculate its AUC for the current validation step. Thus, each model is tested uniquely on data that have not influenced its training in any way. The mean and standard deviation of the 6 AUC values obtained from this cross-validation are reported below as the final AUC and its uncertainty.

The siNet and MIL models are trained in batches of 4,096 6-second frames and in batches of 64 recordings, respectively. In the MIL approach, only 64 randomly-sampled frames of each given recording are included in the batch due to GPU memory limitations. This can be viewed as a variant of data augmentation or, more specifically, as a masking technique. If a model is required to correctly classify a recording based on 64 random frames, it will perform even better when all the frames are available during the prediction phase. We calculate the training AUC using only those 64 frames, whereas the validation and test metrics are always evaluated on entire recordings.

\begin{figure}[!htb]
\centering
\includegraphics[width=0.475\textwidth]{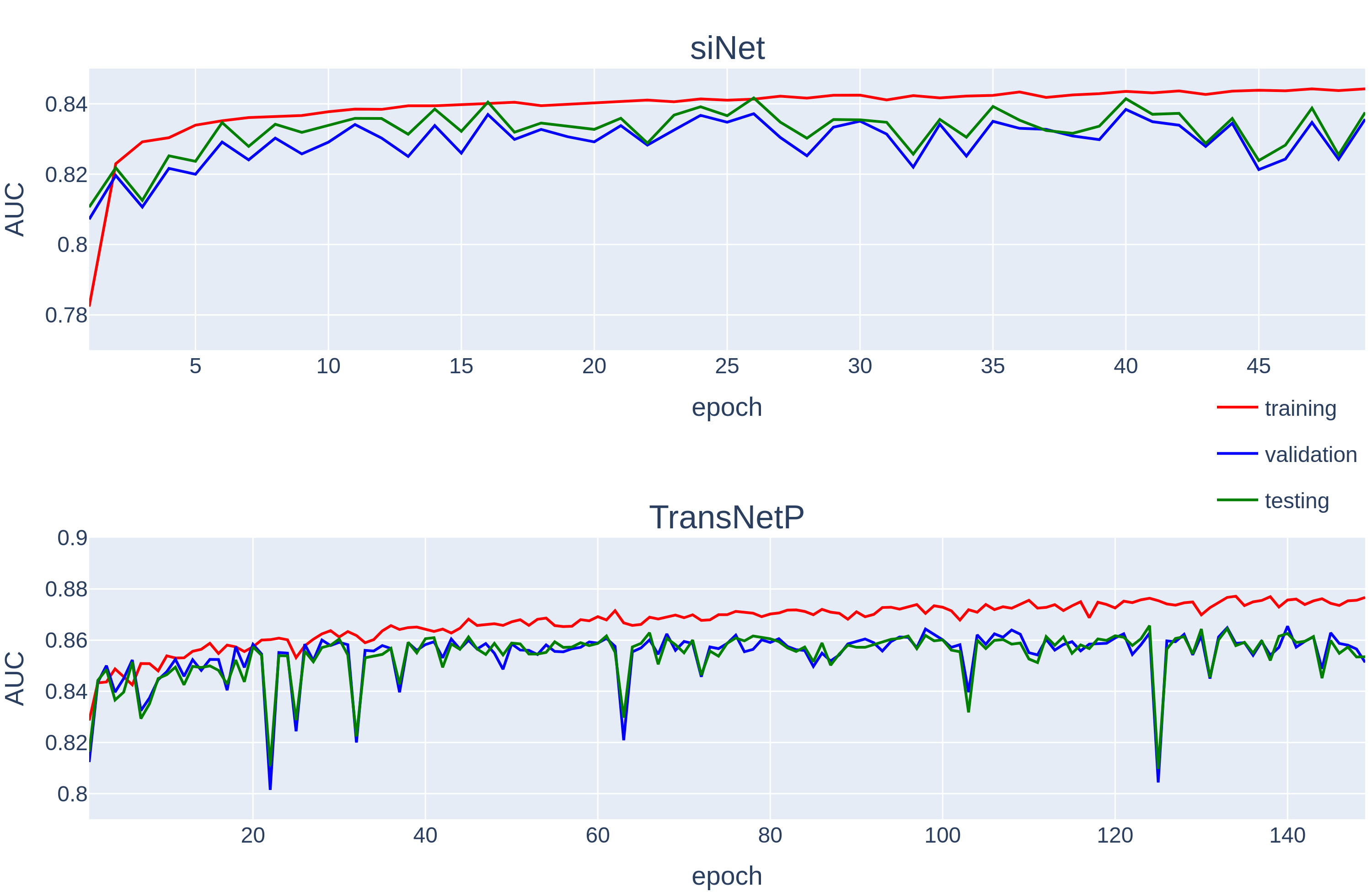}
\caption{Learning curves for the siNet and TransNetP models for the main training phase in a sample cross-validation step.}
\label{fig:learning}
\end{figure}

Sample learning curves are displayed in Fig. \ref{fig:learning}. They show the AUC scores versus epoch numbers for the main training of the siNet and TransNetP models in an exemplary cross-validation step. AUC values on the test fold, validation fold, and four training folds are plotted. The curves are fairly standard, except that the validation and test values for TransNetP suffer from downward spikes at certain epochs. This undesired effect occurs for some MIL models to a greater or lesser extent. We address it by selecting the epoch that yields the highest validation AUC, as described above. This is one reason why we cannot simply stop the training after an a priori fixed number of epochs but must introduce the validation fold.

The classical models are trained and evaluated in cross-validation on the same 6 folds used for the neural networks. The GBE is fitted to the 4 training folds, while the validation fold is used to select the optimal number of trees that maximizes the validation AUC score. The {\tt scikit-learn} classes for SVC and random-forest have no similar functionality, so the validation fold is skipped for RF and RG, and their training terminates based on the training set, as implemented in the library. However, the final AUC scores are always evaluated on the test fold to avoid data leakage and to make the results fully comparable with those obtained for the neural networks.

\subsection{Statistical tests}
\label{sect:test}

The 6-fold cross-validation procedure yields a statistical sample of 6 AUC values for each model and dataset. To compare these results across different samples, we calculate the mean AUC and its standard deviation for each sample. Statistical testing would certainly be desirable to better assess the differences in AUC both among various models within a single dataset and across different datasets for a given model. However, a review devoted to comparing many models on many datasets \citep{demsar2006statistical} indicates that no existing statistical test fully meets our needs due to partially violated assumptions. Keeping this in mind, we nevertheless perform a relatively robust test described below, but we consider its outcomes only as auxiliary indications. Fortunately, these results support our conclusions drawn from the mean AUC values, reinforcing our belief that the presented findings are valid.

We resort to the non-parametric Kruskal-Wallis test \citep{kruskal1952use} and feed it samples of 6 AUC values from cross-validation. We pass samples either from our 11 models tested on a given dataset or from a specific model evaluated on our 7 datasets. The test checks the null hypothesis that samples do not exhibit statistically significant differences in AUC values. If the hypothesis is rejected, the post-hoc Conover-Iman test \citep{conover1979multiple} provides a further pairwise comparison between the samples. We adjust the resulting p-values according to the false discovery rate (FDR) method \citep{FDR}. The lower the p-value, the more certain the statistical difference between the corresponding two samples, and critical p-values of 0.05 or 0.01 are commonly adopted as rejection thresholds.

\section{Results and discussion}

\subsection{Reproduction of the literature results}

As a sanity check, we start by reproducing the relevant TUH results reported by \citet{Gemein2020}. We carry out 6-fold cross-validations for the RF, RG, and siNet models, which are common to both papers (see Section \ref{sect:models}). Because Gemein reports ACC, we also calculate ACC by assuming the standard probability threshold of 0.5. Table \ref{tab:AUC_all} compares the results, and the agreement is well within one standard error.

\begin{table}
\caption{Comparison of cross-validation ACC values for the TUH database obtained in the present study and those reported by \citet{Gemein2020} for the three common models. The equivalent of our siNet is referred to as BD-EEGNet in the original paper.\label{tab:AUC_all}}
\centering
\begin{tabular}{lcc}
\toprule
      & Present study    & \citet{Gemein2020} \\
\midrule
RF    & 82.40 $\pm$ 0.56 & 83.10 $\pm$ 2.44   \\
RG    & 81.25 $\pm$ 1.00 & 81.26 $\pm$ 0.19   \\
siNet & 82.54 $\pm$ 1.64 & 83.36 $\pm$ 1.94   \\
\bottomrule
\end{tabular}
\end{table}

\subsection{Comparison of cross-validation results}
\label{sect:result}

\begin{figure*}[!htb]
\centering
\includegraphics[width=0.75\textwidth]{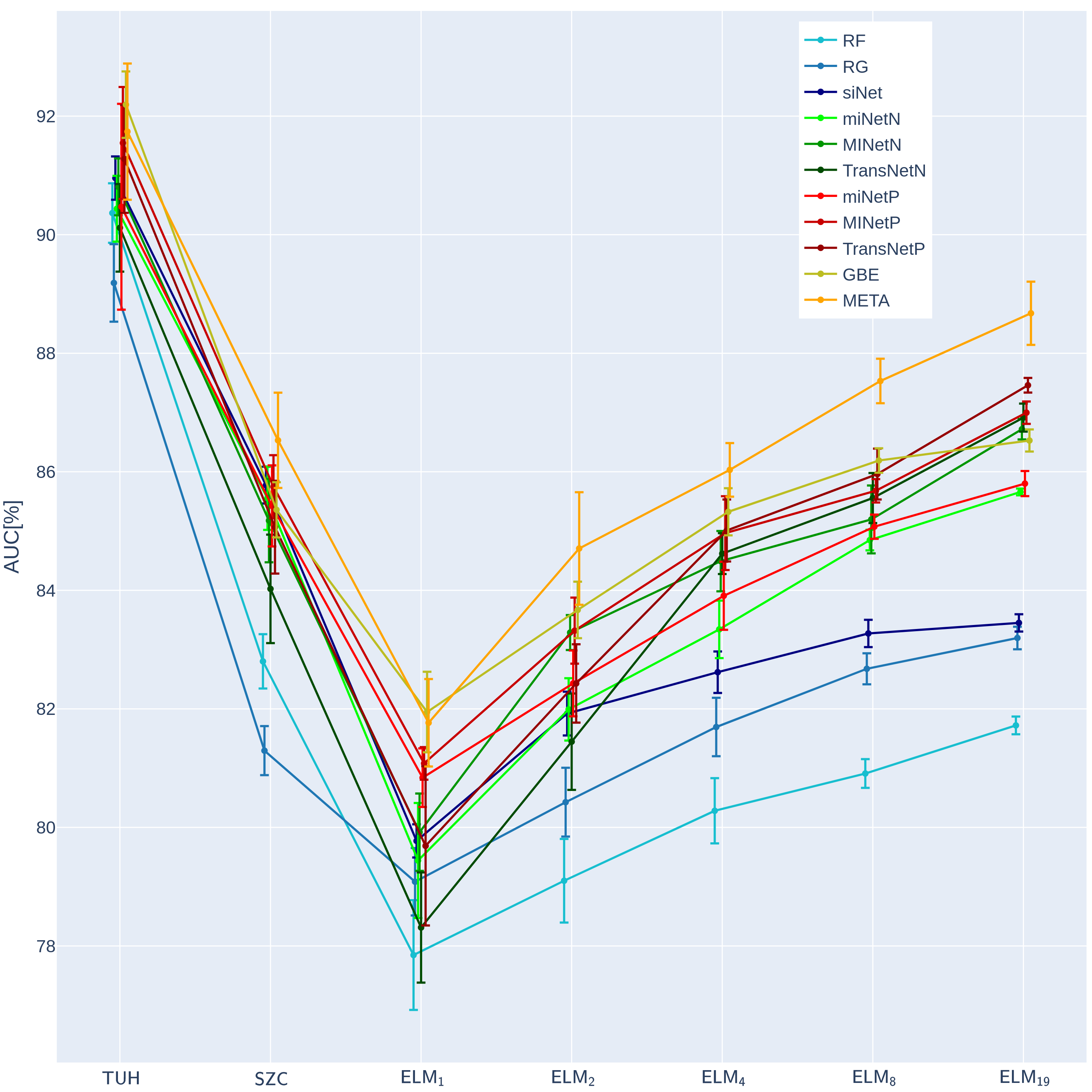} 
\caption{Cross-validation AUC results for our models and datasets. Data points represent mean AUC scores, and error bars indicate the standard error of the mean. Values for reference models inspired by \citet{Gemein2020} are plotted in shades of blue; the MIL models with a naive encoder are in shades of green; and the MIL models with a pretrained encoder are in shades of red.}
\label{fig:AUC_CMP}
\end{figure*}

The cross-validation AUC scores for all our 11 models tested on all 7 considered data subsets are listed in Table \ref{tab:AUC} of \ref{app:B} and presented graphically in Fig. \ref{fig:AUC_CMP}. Results of the statistical tests described in Section \ref{sect:test} are given in \ref{app:C}. Fig. \ref{fig:withinDB} compares different models within each dataset, while Figs. \ref{fig:betweenDB1} and \ref{fig:betweenDB2} compare each model between different datasets. Among other quantities, the figures provide the p-values from the Conover-Iman test, which indicate the pairwise statistical differences between models or datasets.

Several interesting conclusions can be drawn from Fig. \ref{fig:AUC_CMP}. Most strikingly, the performance of every model degrades rapidly across the TUH, SZC, and ELM$_1$ subsets, despite their equal sizes. The statistical significance of the differences in AUC values is supported by the low Conover-Iman p-values obtained for all our models in comparisons between TUH and SZC, as well as between SZC and ELM$_1$ (see Figs. \ref{fig:betweenDB1} and \ref{fig:betweenDB2}). The drop from TUH to SZC can be attributed to the increase in the diversity of pathological conditions. For example, psychogenic seizures often lack the clear electrographic markers typical of epileptic seizures \citep{naganur2019utility}. The decrease in AUC from SZC to ELM$_1$ is apparently due to the inclusion of multiple hospitals. Data from different institutions most likely occupy overlapping but distinct regions of the feature space. If the same amount of data is distributed over many different regions, then each region contains remarkably fewer samples. These points illustrate how different kinds of heterogeneity make EEG classification more challenging and diminish the model's capabilities.

Recent EEG studies \citep{kiessner2023extended, western2021automatic} have confirmed the rather obvious expectation that using larger datasets can improve the quality of pathology detection. Indeed, the AUC values steadily increase through the ELM$_{19}$ subsets due to the increase in data quantity. This also shows that a few thousand recordings present in ELM$_1$ (like in TUH) are not enough to effectively train models on heterogeneous data. At least tens of thousands are required, as in the full ELM$_{19}$. Fig. \ref{fig:AUC_CMP} strongly suggests that the AUC would continue to increase with larger amounts of data, indicating that even more data are actually needed. In Section \ref{sect:asymptote}, we estimate the asymptotic behavior for infinite data.

By using the statistical test of signs, \citet{Gemein2020} conclude that the performance of all their methods on the TUH dataset is statistically equivalent. The Kruskal-Wallis test indicates that the same is true for our models, as seen in the high p-values in Fig. \ref{fig:withinDB} a). One may, therefore, argue that virtually all reasonable architectures perform relatively well on small and homogeneous data, so such data do not truly allow us to identify which approaches are best suited for EEG classification. \citet{kiessner2024reaching} demonstrate that differences between models emerge only with increasing amounts of homogeneous data, while we observe similar behavior for the heterogeneous case. Throughout the ELM$_{19}$ subsets, more advanced models take over while simpler ones lag behind, which is clear from Fig. \ref{fig:AUC_CMP}. The Conover-Iman p-values marked in Fig. \ref{fig:withinDB} c)--g) also illustrate the gradual onset of statistically significant differences between models as the data quantity increases.

We compare our models mostly on the data subsets from ELM$_1$ through ELM$_{19}$ because they represent the fully heterogeneous case. As far as classical methods are concerned, they can be ordered from RF to GBE in terms of the achieved AUC values. Surprisingly, RF performs worse than RG, even though it uses 2,660 frequency-domain features compared to only 190 time-domain ones passed to RG. This may be because the frequency features are normalized, so they do not carry information on the total signal power or amplitude, while the time features are not normalized and do contain that information. Alternatively, the Riemannian projections of the time-domain covariances may indeed be particularly well suited for EEG analysis \citep{wilson2022deep, tibermacine2024riemannian}.

Nevertheless, our GBE model significantly outperforms both RG and RF, which is confirmed by low Conover-Iman p-values between RF and GBE as well as between RG and GBE for all our datasets apart from TUH, as displayed in Fig. \ref{fig:withinDB}. The GBE is likely superior because it uses all the handcrafted features, thus gaining more information about the signal, including the total power. Also, the modern gradient-boosting algorithm is generally considered more powerful, and finally, the model makes use of an ensemble that improves its generalization capabilities.

Let us now proceed to the neural networks. The simple siNet clearly loses the race against the MIL models. This is likely because its single-instance training suffers from the inherent flaw of assigning the label from the entire recording to every single frame, as pointed out in Section \ref{sect:net}. Any architecture based on single-instance learning would exhibit the same weakness, regardless of whether its frame encoder is EEGNet, Deep4, or another. A vast majority of neural EEG models considered so far fall into this category.

Even the simplest MIL model, miNet, clearly outperforms siNet, despite both having the same architecture and differing only in the training routine, as discussed in Section \ref{sect:net}. This demonstrates the significance of training neural models on entire recordings, to which the target labels pertain. Including the attention mechanism in the MINet architecture further boosts performance. Apparently, the appropriate selection of frames that carry the most information is quite important in EEG analysis. Accounting for the relations between the frames through self-attention results in even higher AUC values, and the resulting TransNetP performs best among our single models. This observation is supported by the low Conover-Iman p-values between TransNetP and almost all other single models in a pairwise comparison on ELM$_{19}$, as given in Fig. \ref{fig:withinDB} g). The superiority of TransNetP aligns with the overall success of the transformer architecture in a wide range of applications.

Networks with a pretrained encoder consistently achieve better results than those trained entirely from scratch. This improvement is modest but still meaningful. Although training on full recordings is methodologically more accurate (refer to Section \ref{sect:net}), it imposes fewer constraints on the encoder because only one target label must be reproduced while many frames are available. Pretraining the encoder on single frames confines its behavior more strongly because every frame has a separate label. Although assigning the overall label of a recording to every frame is not fully appropriate, this pretraining strategy is effective enough to guide the encoder in the right direction for subsequent MIL training (see Section \ref{sect:net}).

\begin{figure*}[!htb]
\centering
\includegraphics[width=0.75\textwidth]{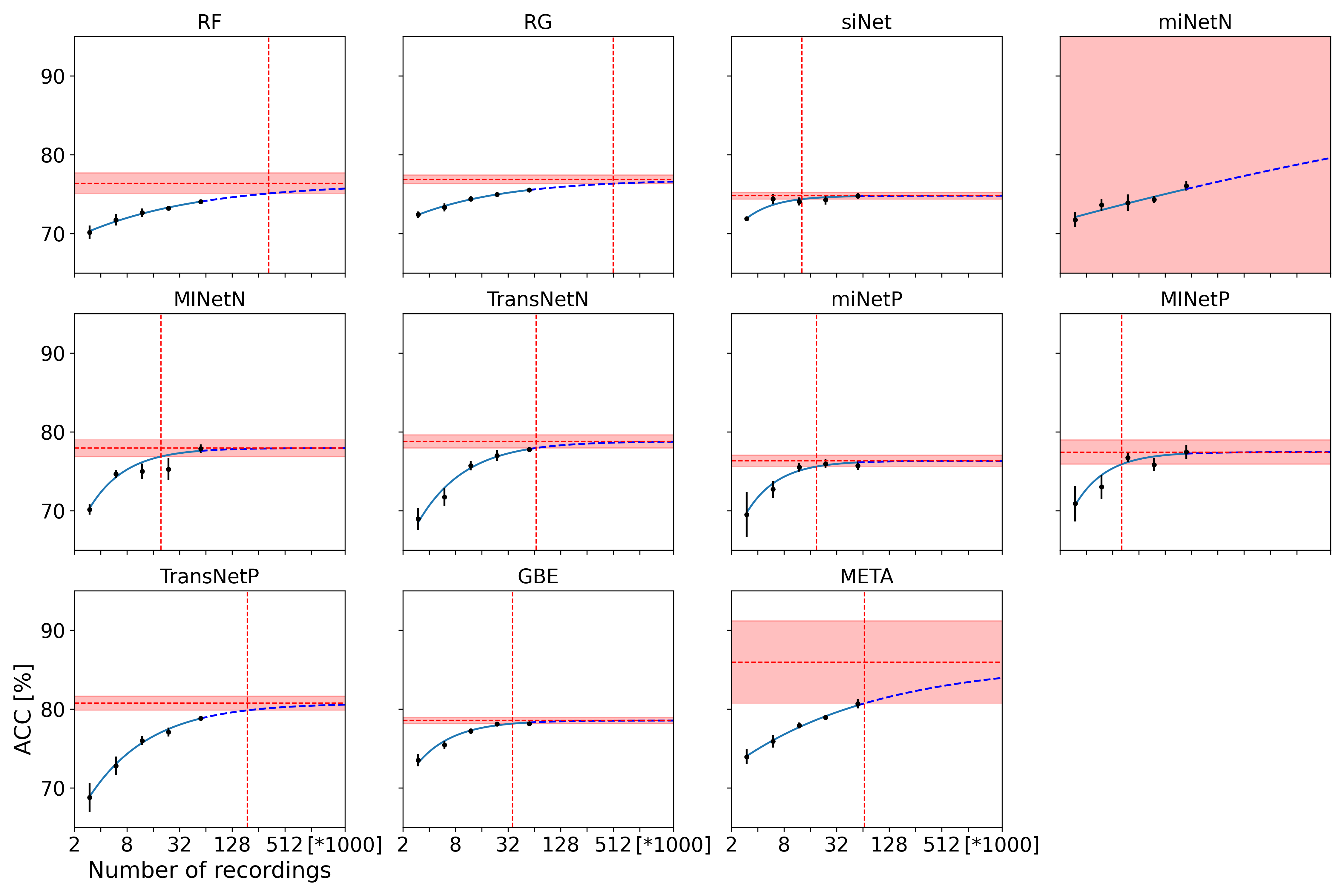}
\caption{Visualization of the saturation power law (\ref{eq:pow-law-ACC}) fitted to ACC[\%] values for all models. The solid blue line represents the fit to the available data, and the dashed line indicates extrapolation. The horizontal red line shows the estimated asymptote, and the shaded region marks the error corridor. The vertical line indicates the number of recordings, $N_{DB}$, for which the estimated ACC approaches the asymptotic value of ACC$_{\infty}$ within one standard error.}
\label{fig:fits_ACC}
\end{figure*}

As expected, the meta-model composed of TransNetP, MINetP, and GBE indisputably outperforms all the single models. The statistical differences in the corresponding AUC scores are confirmed by the low p-values of the Conover-Iman tests shown in Fig. \ref{fig:withinDB}, particularly for the largest ELM$_{19}$ subsets. The meta-model takes into account both the handcrafted features and the ones extracted by the neural networks, thus getting more information about the signal at hand. It may also benefit from an implicit regularization effect. By averaging the decision surfaces of its component models, it yields a smoother decision boundary, which is less prone to overfitting. Fig. \ref{fig:betweenDB2} k) reports a rather low Conover-Iman p-value for the meta-model in comparison between TUH and ELM$_{19}$, which is a preliminary indication that the model might perform similarly on large heterogeneous data as it does for small and homogeneous ones. More arguments in favor of such a conclusion are given in Section \ref{sect:asymptote}.

The AUC curve for our best classical model, GBE, starts to saturate for the largest ELM$_{19}$ subsets. Our conjecture is that the model simply cannot extract more information from the limited number of handcrafted features. It cannot create new features either. In contrast, the neural networks extract features themselves from the raw EEG signal and thus have the full information at their disposal. The networks presumably identify more and more useful properties of the signal as they receive more data. Some of them continue to improve and eventually outperform GBE for the full ELM$_{19}$ set. Indeed, the corresponding differences in the AUC values between GBE and the pretrained networks with (self)-attention (i.e., MINetP and TransNetP) are statistically significant, as illustrated by the low Conover-Iman p-values given in Fig. \ref{fig:withinDB} g).

Finally, note that the AUC curves for the MINet and TransNet models do not exhibit signs of saturation. Consequently, the same is true for the meta-model. This leads us to expect that larger datasets will allow those architectures to attain even higher AUC values. We elaborate on this point in the next section.

\subsection{Asymptotic behavior}
\label{sect:asymptote}

The curves plotted in Fig. \ref{fig:AUC_CMP} strongly suggest that some models would achieve higher AUC values if more data were available. Therefore, it is interesting to estimate the upper AUC limit and the quantity of data needed to reach it. Similar considerations have recently gained significant attention in other ML domains, and a saturation power law is often used to approximate the relationship between accuracy and data size \citep{kiessner2024reaching}. This law can be formulated as follows:
\begin{equation}
\mathrm{ACC}(n) = \mathrm{ACC}_\infty - \alpha n^{-\beta}
\label{eq:pow-law-ACC}
\end{equation}
In our case, $n$ is the number of EEG recordings, while $\alpha$, $\beta$, and $\mathrm{ACC}_\infty$ are parameters that can be fitted to the existing dataset. $\mathrm{ACC}_\infty$ has a clear interpretation as the asymptotic ACC value in the limit of infinite data.

For each of our models, we fitted the saturation power law (\ref{eq:pow-law-ACC}) to the cross-validation ACC values for the ELM$_{19}$ subsets using the standard least-squares procedure. The resulting curves are plotted in Fig. \ref{fig:fits_ACC}, along with the data points. The asymptotic value of ACC and its uncertainty estimated from the fit are marked with a horizontal line and an error corridor. Although the fits are generally reasonable, some data points deviate from the fitted curves, errors on the asymptotic ACC values are sometimes large, and the fit for miNetN did not even converge. Additionally, fits for neural networks are systematically less robust.

\begin{figure*}[!htb]
\centering
\includegraphics[width=0.75\textwidth]{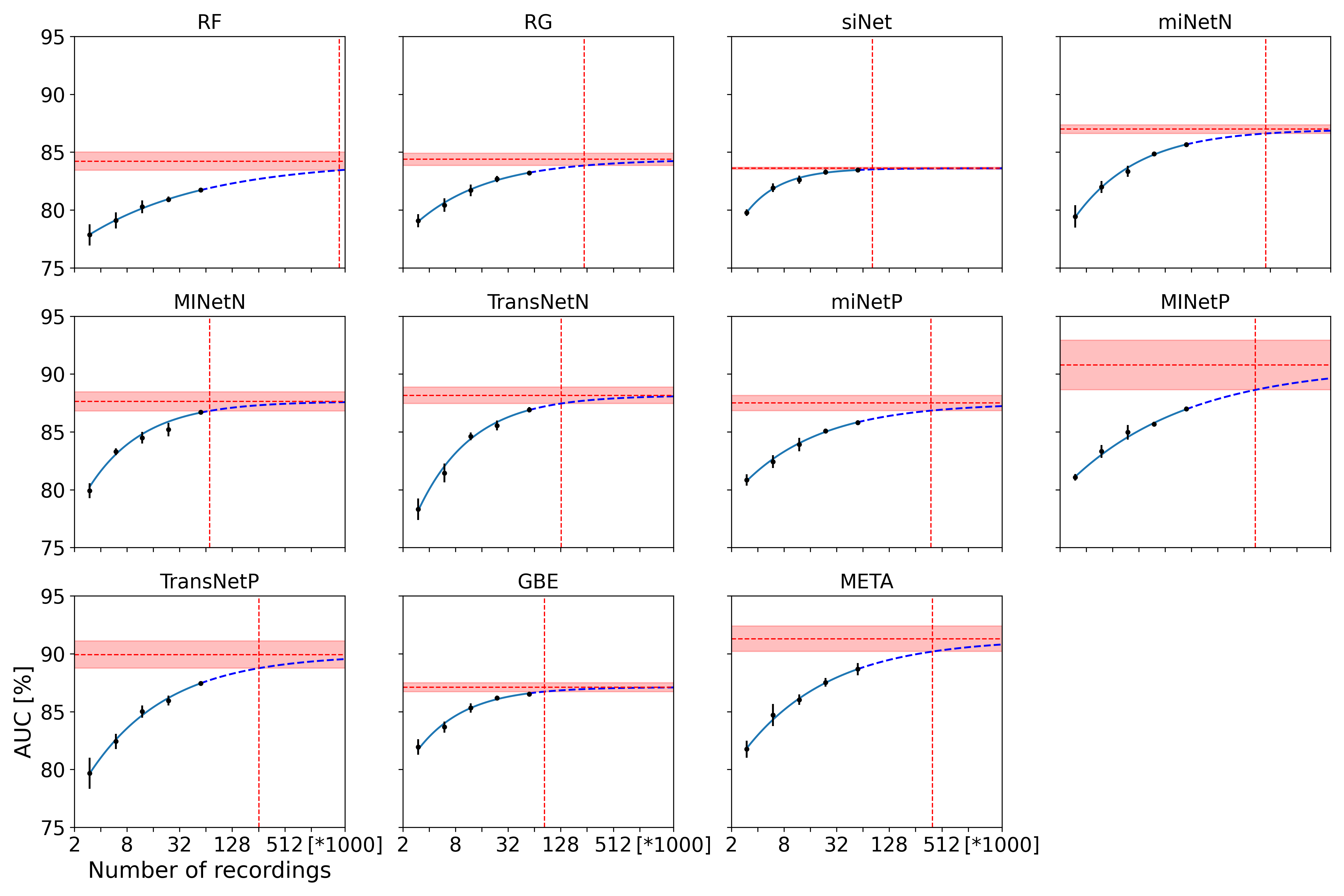}
\caption{Visualization of the  saturation power law (\ref{eq:pow-law-ACC}) fitted to AUC[\%] values for all models. The solid blue line represents the fit to the available data, and the dashed line indicates extrapolation. The horizontal red line shows the estimated asymptote, and the shaded region marks the error corridor. The vertical line indicates the number of recordings, $N_{DB}$, for which the estimated AUC approaches the asymptotic value of AUC$_{\infty}$ within one standard error.}
\label{fig:fits_AUC}
\end{figure*}

Our explanation for the latter point is that the optimizers used to train neural networks somewhat chaotically vary the bias term in the last layer that generates the logit values. This is equivalent to chaotically adjusting the threshold on the resulting probability within a certain range. Because the ACC value depends on that threshold, it exhibits redundant oscillations. In contrast, the AUC metric is independent of the probability threshold and, therefore, behaves more smoothly. Given these observations, we also fitted the power law (\ref{eq:pow-law-ACC}) to the cross-validation AUC values, substituting AUC in place of ACC. The results are presented in Fig. \ref{fig:fits_AUC}. The AUC fits indeed demonstrate better agreement with the data and smaller errors on the asymptotic values. The standard $R^2$ score, which measures fit quality, is also closer to the expected value of one, as listed in Table \ref{tab:ACC_fit}.

\begin{table*}[!htb]
\caption{Estimations of the asymptotic ACC and AUC values according to the saturation power law (\ref{eq:pow-law-ACC}), along with the estimated number of recordings $N_{DB}$ (in thousands) needed for each model to approach the asymptotic value within one standard error. Goodness of fit is measured by $R^2$. \label{tab:ACC_fit}}
\centering
\begin{tabular}{lccrccr}
\toprule
          & \multicolumn{3}{c}{ACC}                           & \multicolumn{3}{c}{AUC}                           \\
\midrule
          & $\mathbf{ACC_{\infty}}[\%]$ & $R^2$ & $N_{DB}$[k] & $\mathbf{AUC_{\infty}}[\%]$ & $R^2$ & $N_{DB}$[k] \\
\midrule
RF        &\textbf{76.4 $\pm$ 1.3}      & 0.97  & 336         & \textbf{84.2 $\pm$ 0.8}     & 0.99  & 2,148       \\
RG        &\textbf{76.9 $\pm$ 0.5}      & 0.97  & 510         & \textbf{84.4 $\pm$ 0.5}     & 0.97  & 237         \\
siNet     &\textbf{74.8 $\pm$ 0.4}      & 0.60  & 13          & \textbf{83.6 $\pm$ 0.1}     & 0.96  & 82          \\
miNetN    & \multicolumn{3}{c}{fit did not converge}          & \textbf{87.0 $\pm$ 0.4}     & 0.98  & 455         \\
MINetN    & \textbf{78.0 $\pm$ 1.1}     & 0.87  & 20          & \textbf{87.6 $\pm$ 0.8}     & 0.89  & 71          \\
TransNetN & \textbf{78.8 $\pm$ 0.8}     & 0.97  & 67          & \textbf{88.2 $\pm$ 0.7}     & 0.97  & 130         \\
miNetP    & \textbf{76.3 $\pm$ 0.7}     & 0.94  & 19          & \textbf{87.5 $\pm$ 0.7}     & 0.98  & 383         \\
MINetP    & \textbf{77.4 $\pm$ 1.5}     & 0.88  & 10          & \textbf{90.8 $\pm$ 2.1}     & 0.95  & 346         \\
TransNetP & \textbf{80.8 $\pm$ 0.9}     & 0.99  & 191         & \textbf{89.9 $\pm$ 1.2}     & 0.98  & 258         \\
GBE       & \textbf{78.6 $\pm$ 0.4}     & 0.84  & 36          & \textbf{87.1 $\pm$ 0.4}     & 0.93  & 83          \\
META      & \textbf{86.0 $\pm$ 5.2}     & 0.97  & 66          & \textbf{91.3 $\pm$ 1.1}     & 0.99  & 401         \\
\bottomrule
\end{tabular}
\end{table*}

Apart from $R^2$, Table \ref{tab:ACC_fit} provides the fitted asymptotic values of ACC and AUC. Our best model, the meta-model, could theoretically achieve an AUC score of roughly 91\%, which is about the mean value reached by all our models on the TUH database. \citet{kiessner2024reaching} estimated the asymptotic ACC for increasing quantities of homogeneous data similar to TUH \citep{kiessner2023extended} and arrived at values between 85 and 87\%, depending on the model. This aligns well with the asymptotic ACC of 86\% achieved by our meta-model for ELM$_{19}$. These two observations suggest that the meta-model could perform as well for heterogeneous data as it does for homogeneous data, provided that sufficiently large datasets are available. In other words, large quantities of EEG data may potentially compensate for their diversity. However, it is important to note that the power law is not fully rooted in fundamental principles, and these predictions are, to some extent, speculative.

Table \ref{tab:ACC_fit} also lists the number of recordings, $N_{DB}$, for which the fitted ACC and AUC scores approach their asymptotic limits within one standard error in ACC$_\infty$ and AUC$_\infty$, respectively. These limits are marked in Figs. \ref{fig:fits_ACC} and \ref{fig:fits_AUC} by vertical lines. $N_{DB}$ represents the data size needed for the models to attain their asymptotic limits, given that the limits are known only up to a certain error. Although such a measure is arbitrary and comes with a significant uncertainties, we believe that $N_{DB}$ correctly estimates the order of magnitude. From the more reliable AUC data in Table \ref{tab:ACC_fit}, we infer that at least tens, and likely hundreds of thousands, of recordings are needed for the best models to approach their full capabilities. One may also speculate that collecting even more data would yield only slow and negligible improvements due to the power character of the saturation law (\ref{eq:pow-law-ACC}), given the logarithmic scale for the abscissa in Figs. \ref{fig:fits_ACC} and \ref{fig:fits_AUC}. However, the presented figures may change if more heterogeneity is introduced by including data from several countries etc.

\section{Limitations and outlook}

A major limitation of our study is that a significant portion, or even a majority, of patients with EEG pathologies are under pharmacological treatment. It is, therefore, possible that the models are detecting EEG patterns induced by medication rather than the pathology itself. Unfortunately, we are unable to assess the real impact of this potential confusion, because there is no precise and reliable information about medication use in our database. We suspect that other studies on EEG pathology detection also suffer from the same uncertainty. Further research should address this issue by creating longitudinal databases that contain several types of recordings, such as abnormal recordings collected before pharmacological treatment is administered, abnormal recordings collected during treatment, normal recordings collected toward the end of treatment when the pathological symptoms have already disappeared, normal recordings collected after treatment has been completed, and normal recordings from healthy patients who have never taken relevant medication.

Furthermore, the ELM$_{19}$ data were collected only in one country. An international effort could provide significantly larger datasets, making them more representative and more heterogeneous.

Our results indicate that as the quantity of data increases, the best models can still improve their performance and may even reach the asymptotic AUC limit of about 91\%. This raises the question of where that value comes from and whether it can be surpassed. Like \citet{kiessner2024reaching}, we believe that the limit is mostly due to moderate inter-rater agreement. Opposite labeling of very similar recordings causes the decision boundary to become spuriously curved or ragged, leading to overfitting and deteriorated performance. In simple terms, an ML model cannot reconcile contradictory human ratings.

A possible remedy is the use of fuzzy labels. In this approach, each recording is annotated by several independent experts as 0 for normal or 1 for pathological. These ratings are then averaged to yield a fuzzy label, which estimates the probability that a randomly chosen rater annotates the recording as pathological. An ML model is then trained to reproduce these probabilities. Averaged ratings are much less contradictory than individual ones, enabling the model to fit them more accurately. This procedure can be viewed as a data-driven regularization that makes the decision boundary smoother and more fuzzy, which in turn reduces overfitting and better elucidates the actual patterns hidden in the data. The manifold mixup proposed by \citet{verma2019manifold} has a similar effect. The output of the resulting model naturally measures the probability or degree of pathology. Even though such a measure is not perfect, it is likely the best that can be derived from human ratings because it reflects the inter-subjective knowledge of many experts. A robustly estimated degree of pathology can be more useful for clinicians than a noisy categorical prediction. This analysis requires data labeled by many raters, but it would be sufficient to assign fuzzy labels only to the most doubtful recordings situated close to the decision boundary.

An even more ambitious solution would be to train models on EEG recordings labeled based on a full clinical examination rather than the EEG signal alone. If such training succeeds, the resulting models could potentially recognize EEG patterns that are not visible to the naked eye, at least for some pathological disorders. Unfortunately, we are not aware of any clinical database large enough to attempt ML training.

\section{Conclusions}

In our setting, data heterogeneity makes EEG classification significantly more difficult and degrades the performance of ML models. Heterogeneity arises from collecting data across multiple hospitals and the diversity of pathological conditions, which are not limited to epileptic disorders.

The performance of ML models improves with increasing quantities of data. In the asymptotic limit of infinite data volume, the best models may theoretically achieve AUC scores of approximately 91\%, which is comparable to the results for homogeneous datasets. This suggests that very large amounts of data can eventually compensate for data heterogeneity, justifying the need to collect even larger corpora of EEG recordings, comprising tens or even hundreds of thousands of samples.

Virtually all relevant ML models perform comparably well on small, homogeneous datasets, while significant differences emerge only with increasing data size. Therefore, only large, heterogeneous databases may help identify architectures best suited for EEG classification in general.

Regarding classical models, the best results are obtained by an ensemble of gradient-boosted classifiers that utilize both time-domain and frequency-domain handcrafted features, including information about signal amplitude. However, classical models cannot extract more information from the signal than is contained in the handcrafted features, which inherently limits their capabilities, particularly for large datasets.

In the case of neural networks, it seems that training on entire recordings within an MIL paradigm, as well as including attention and self-attention mechanisms, is essential. Consequently, the transformer architecture prevails among both neural and classical models. Pretraining the encoder on individual frames also helps improve performance.

Because neural networks and classical models presumably exploit different properties of the EEG signal, even more information can be extracted by merging the two types into a meta-model. Indeed, our variant of the meta-model performs best among all the methods that we tested. It achieves an AUC value exceeding 88\% on our largest heterogeneous dataset of 55,787 recordings and is expected to reach 91\% in the asymptotic limit of infinite data.

The use of medication by most patients with EEG pathologies poses a challenge for automatic EEG classification and should be addressed through further research. Collecting EEG annotations from multiple raters to form fuzzy labels could potentially enable ML models to overcome current asymptotic limits on predictive accuracy.

\section*{Declarations}

\paragraph{Declaration of generative AI in scientific writing}
The authors used no AI assistance when writing the text of this article.

\paragraph{Data availability}
The TUH database can be obtained from The Neural Engineering Data Consortium at \url{https://isip.piconepress.com/projects/nedc/html/tuh_eeg/#c_tuab}. The ELM$_{19}$ database is available from Elmiko Biosignals sp. z o.o. upon reasonable request to \url{EEG.database@elmiko.pl}.

\paragraph{Conflict of interest}
The authors declare that they have no conflicts of interest.

\paragraph{Ethics approval}
All procedures involving human participants were performed in accordance with the ethical standards of the institutional and/or national research committee, as well as the 1964 Helsinki Declaration and its later amendments or comparable ethical standards.

\paragraph{Acknowledgments}
This study was supported by The National Centre for Research and Development (NCBR) grant number POIR.01.01.01-00-1074/21 and partially supported by the Center for Machine Learning, University of Warsaw (IDUB, Action I.3.7).

\paragraph{CRediT authorship contribution statement}
Conceptualization: M.P., M.D., Pr.O., P.N., Pa.O., J.R., Ja.Z.; Creation and anonymization of the ELM$_{19}$ dataset: P.N., Pa.O., Jo.Z.; Data curation: M.P., M.D., Pa.O.; Analysis and interpretation of results: M.P., M.D., Pr.O., J.R., Ja.Z.; Drafting the manuscript: M.P., M.D., Pr.O., J.R., Ja.Z.

\bibliographystyle{elsarticle-num-names}

\onecolumn 

\appendix

\section{Color code for neural-network architectures}

\begin{figure}[!htb]
\centering
\includegraphics{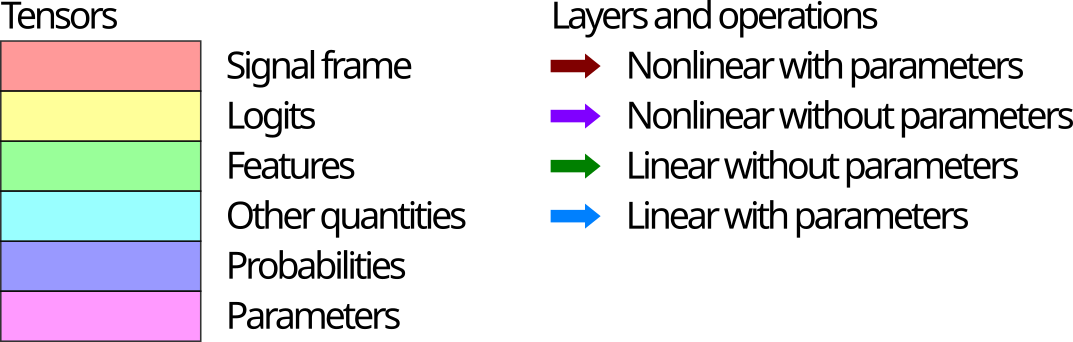}
\caption{Color code used throughout the paper to visualize neural-network architectures.}
\label{fig:colors}
\end{figure}

\section{AUC values for the cross-validation results}
\label{app:B}

\begin{table*}[!htb]
\caption{Mean cross-validation AUC values and their standard errors obtained for each model and dataset. These results are visualized in Fig. \ref{fig:AUC_CMP} \label{tab:AUC}}
\centering
\begin{tabular}{llllllll}
\toprule
          & $\mathrm{TUH}$ & $\mathrm{SZC}$ & $\mathrm{ELM}_1$ & $\mathrm{ELM}_2$ & $\mathrm{ELM}_4$ & $\mathrm{ELM}_8$ & $\mathrm{ELM}_{19}$ \\
\midrule
RF        & 90.4 $\pm$ 0.5 & 82.8 $\pm$ 0.5 & 77.8 $\pm$ 0.9   & 79.1 $\pm$ 0.7   & 80.3 $\pm$ 0.6   & 80.9 $\pm$ 0.2   & 81.7 $\pm$ 0.1      \\
RG        & 89.2 $\pm$ 0.7 & 81.3 $\pm$ 0.4 & 79.1 $\pm$ 0.6   & 80.4 $\pm$ 0.6   & 81.7 $\pm$ 0.5   & 82.7 $\pm$ 0.3   & 83.2 $\pm$ 0.2      \\
siNet     & 91.0 $\pm$ 0.4 & 85.8 $\pm$ 0.3 & 79.8 $\pm$ 0.3   & 81.9 $\pm$ 0.4   & 82.6 $\pm$ 0.4   & 83.3 $\pm$ 0.2   & 83.5 $\pm$ 0.1      \\
miNetN    & 90.4 $\pm$ 0.6 & 85.6 $\pm$ 0.5 & 79.4 $\pm$ 1.0   & 82.0 $\pm$ 0.5   & 83.3 $\pm$ 0.5   & 84.8 $\pm$ 0.2   & 85.7 $\pm$ 0.1      \\
MINetN    & 90.8 $\pm$ 0.5 & 85.2 $\pm$ 0.7 & 79.9 $\pm$ 0.7   & 83.3 $\pm$ 0.3   & 84.5 $\pm$ 0.5   & 85.2 $\pm$ 0.6   & 86.7 $\pm$ 0.2      \\
TransNetN & 90.1 $\pm$ 0.7 & 84.0 $\pm$ 0.9 & 78.3 $\pm$ 0.9   & 81.4 $\pm$ 0.8   & 84.6 $\pm$ 0.3   & 85.6 $\pm$ 0.4   & 86.9 $\pm$ 0.2      \\
miNetP    & 90.5 $\pm$ 1.7 & 85.4 $\pm$ 0.7 & 80.8 $\pm$ 0.5   & 82.4 $\pm$ 0.6   & 83.9 $\pm$ 0.6   & 85.1 $\pm$ 0.2   & 85.8 $\pm$ 0.2      \\
MINetP    & 91.5 $\pm$ 0.9 & 85.7 $\pm$ 0.5 & 81.1 $\pm$ 0.3   & 83.3 $\pm$ 0.6   & 85.0 $\pm$ 0.6   & 85.7 $\pm$ 0.2   & 87.0 $\pm$ 0.2      \\
TransNetP & 91.3 $\pm$ 0.9 & 85.1 $\pm$ 0.8 & 79.7 $\pm$ 1.3   & 82.4 $\pm$ 0.7   & 85.0 $\pm$ 0.5   & 86.0 $\pm$ 0.4   & 87.5 $\pm$ 0.1      \\
GBE       & 92.2 $\pm$ 0.6 & 85.4 $\pm$ 0.5 & 81.9 $\pm$ 0.7   & 83.7 $\pm$ 0.5   & 85.3 $\pm$ 0.4   & 86.2 $\pm$ 0.2   & 86.5 $\pm$ 0.2      \\
META      & 91.7 $\pm$ 1.1 & 86.5 $\pm$ 0.8 & 81.8 $\pm$ 0.7   & 84.7 $\pm$ 0.9   & 86.0 $\pm$ 0.5   & 87.5 $\pm$ 0.4   & 88.7 $\pm$ 0.5      \\
\bottomrule
\end{tabular}
\end{table*}

\newpage

\section{Statistical tests}
\label{app:C}

\begin{figure*}[!htb]
\centering
\includegraphics[width=0.85\textwidth]{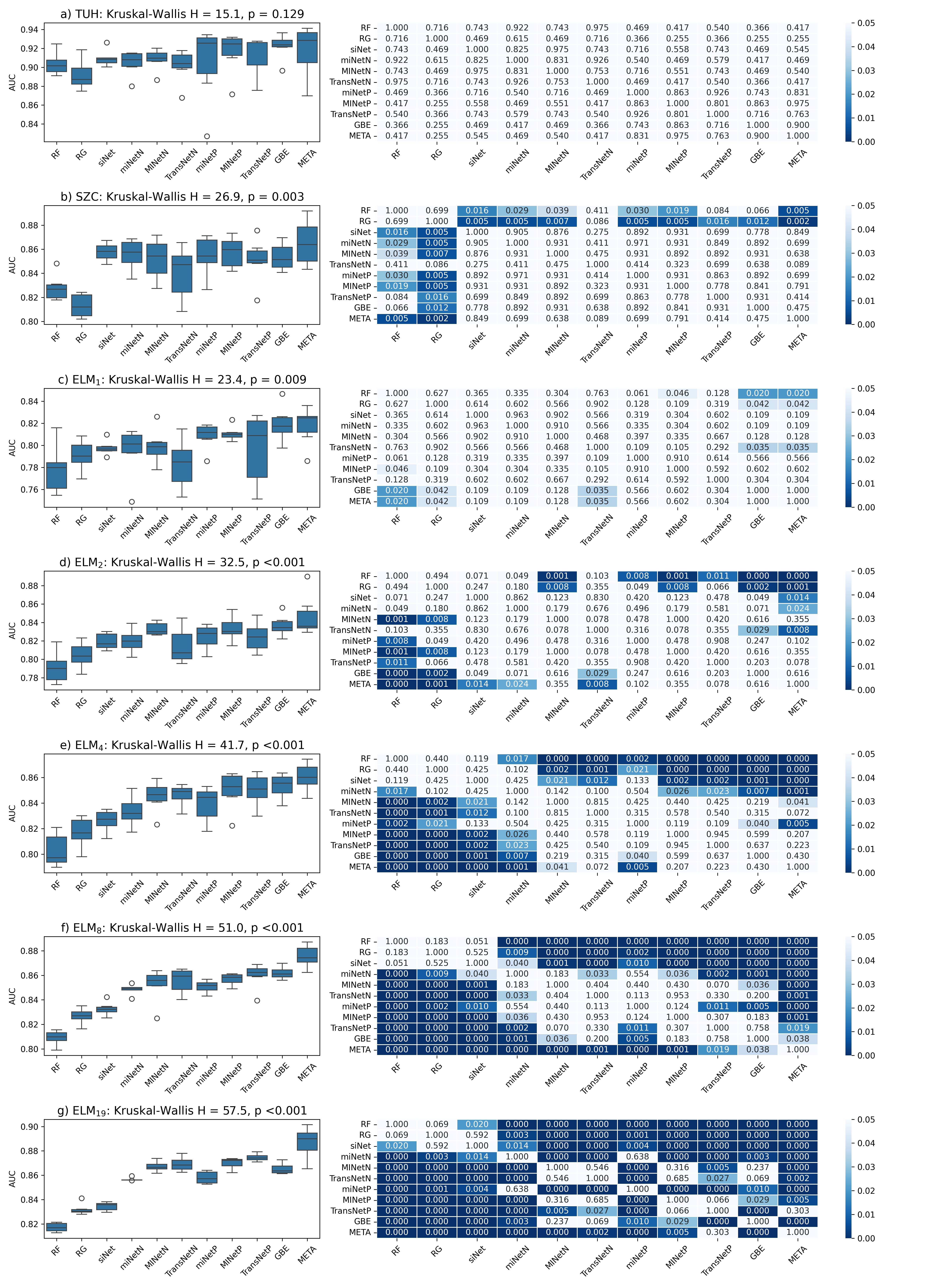}
\caption{Statistical comparison of AUC scores yielded by the models within each dataset. Left -- boxplots and Kruskal-Wallis test statistic H and p-value; right -- post-hoc Conover-Iman p-values adjusted using the FDR method.}
\label{fig:withinDB}
\end{figure*}

\begin{figure*}[!htb]
\centering
\includegraphics[width=0.85\textwidth]{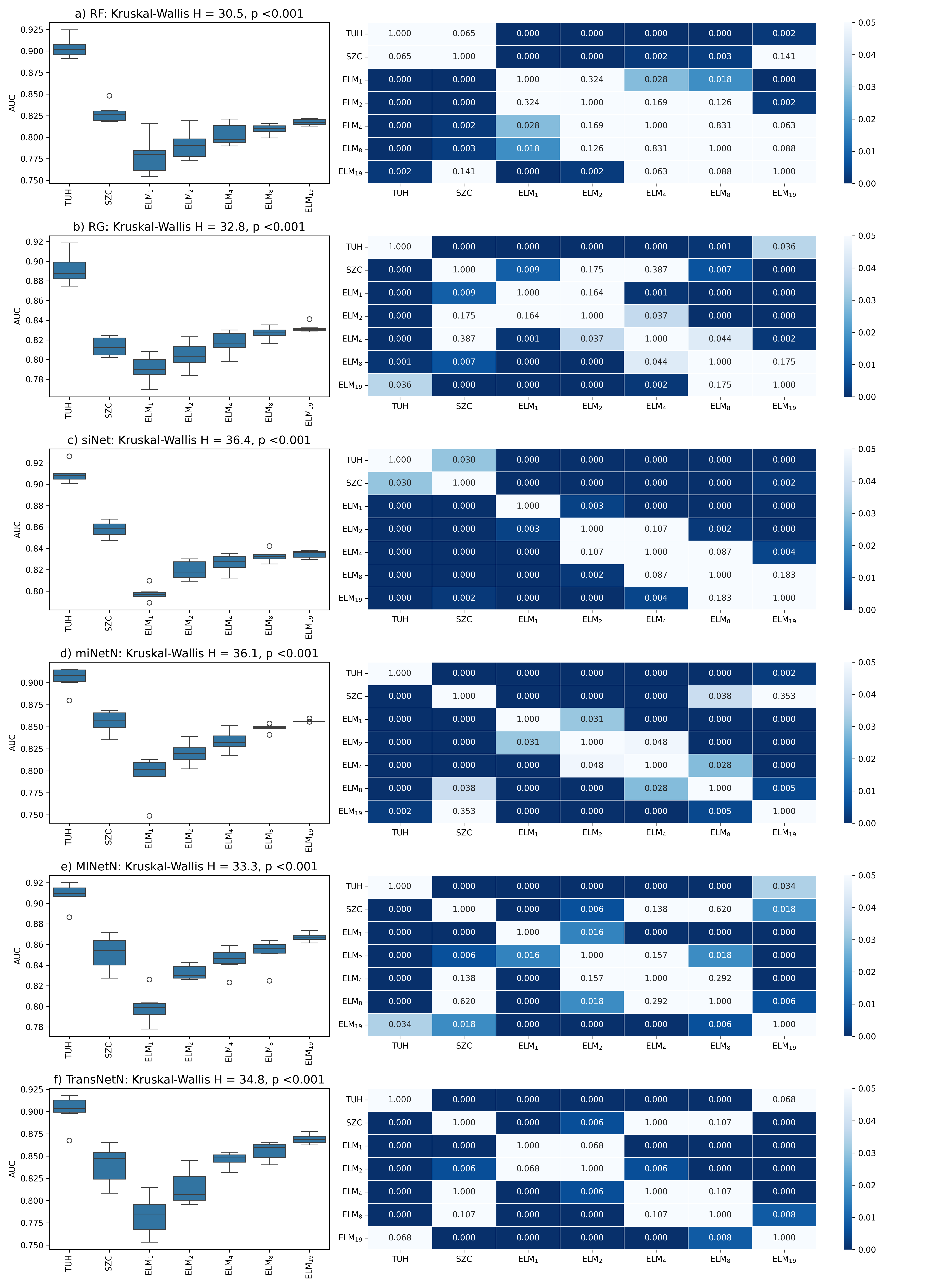}
\caption{Statistical comparison of AUC scores yielded by each model across different datasets. Left -- boxplots and Kruskal-Wallis test statistic H and p-value; right -- post-hoc Conover-Iman p-values adjusted using the FDR method.}
\label{fig:betweenDB1}
\end{figure*}

\begin{figure*}[!htb]
\centering
\includegraphics[width=0.85\textwidth]{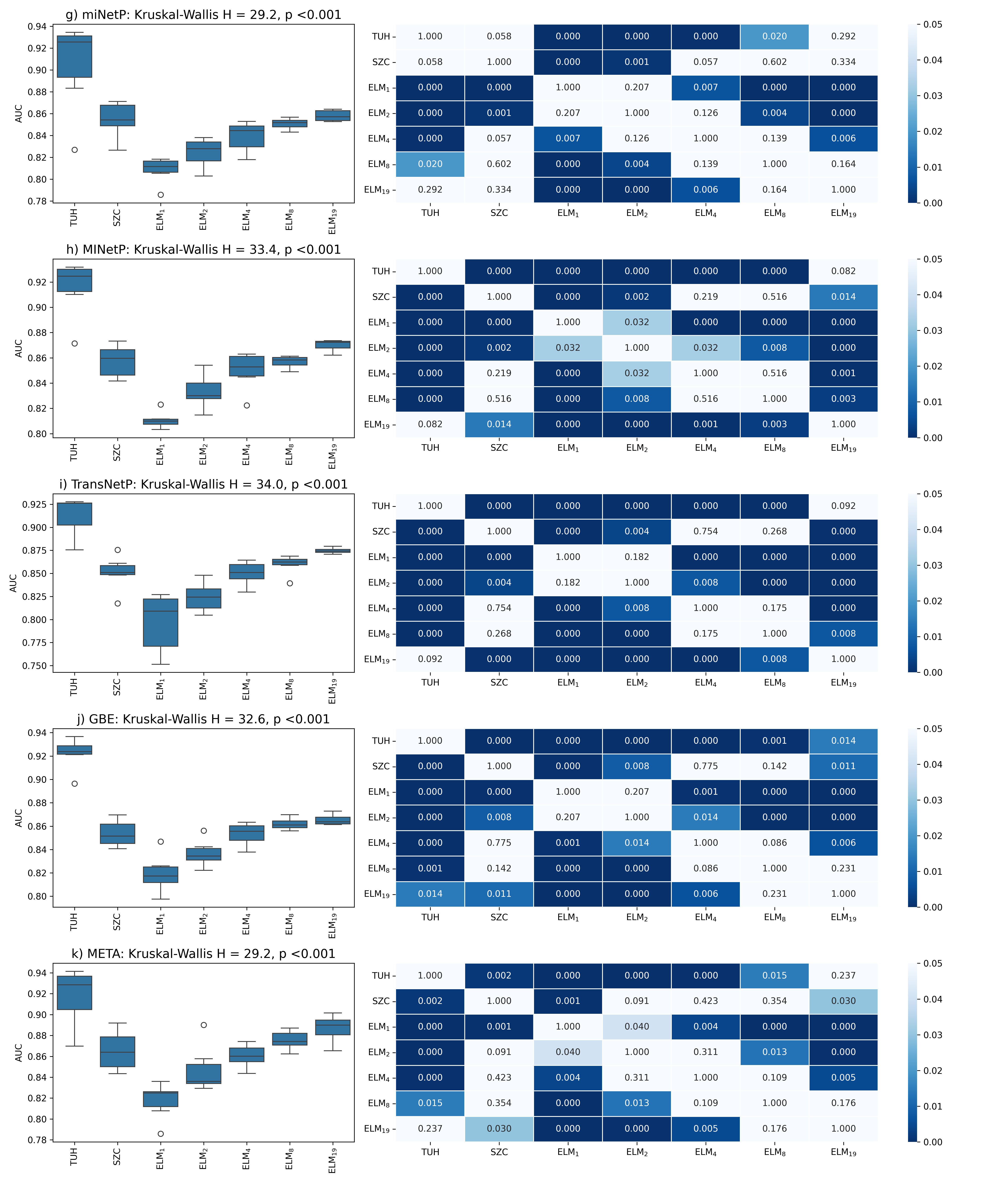}
\caption{Statistical comparison of AUC scores yielded by each model across different datasets. Left -- boxplots and Kruskal-Wallis test statistic H and p-value; right -- post-hoc Conover-Iman p-values adjusted using the FDR method.}
\label{fig:betweenDB2}
\end{figure*}

\end{document}